\documentclass[epss, 10pt]{article}
\input{psfig.sty}
\usepackage{amsmath}
\usepackage{xspace}
\usepackage{verbatim}
\usepackage{latexsym}
\usepackage{amssymb}
\usepackage{bm}
\usepackage[dvips]{graphics}
\usepackage{graphicx}
\usepackage{psfrag}
\begin{document}
\renewcommand{\textfraction}{0}

\title{Universal Filtering via Hidden Markov Modeling
\footnote{Authors are with the department of electrical engineering,
Stanford University, Stanford, CA 94305, USA. e-mails: {\tt
\{tsmoon, tsachy\}@stanford.edu}. The material in this paper was
partially presented at the 2005 ISIT, \cite{Taesup05}.
 The research
was partially supported  by NSF grant CCR-0311633}}
\author{
Taesup Moon  \and Tsachy Weissman }
\maketitle \thispagestyle{empty}
\begin{abstract}
The problem of discrete universal filtering, in which the components
of a discrete signal emitted by an unknown source and corrupted by a
known DMC are to be causally estimated, is considered. A family of
filters are derived, and are shown to be universally asymptotically
optimal in the sense of achieving the optimum filtering performance
when the clean signal is stationary, ergodic, and satisfies an
additional mild positivity condition. Our schemes are comprised of
approximating the noisy signal using a hidden Markov process (HMP)
via maximum-likelihood (ML) estimation, followed by the use of the
forward recursions for HMP state estimation. It is shown that as the
data length increases, and as the number of states in the HMP
approximation increases, our family of filters attain the
performance of the optimal distribution-dependent filter.\\

\end{abstract}

\normalsize

\setlength{\baselineskip}{1.3\baselineskip}
\newtheorem{claim}{Claim}
\newtheorem{guess}{Conjecture}
\newtheorem{defn}{Definition}
\newtheorem{fact}{Fact}
\newtheorem{assumption}{Assumption}
\newtheorem{theorem}{Theorem}
\newtheorem{lem}{Lemma}
\newtheorem{cor}{Corollary}
\newtheorem{proof}{Proof}
\newtheorem{pfth}{Proof of Theorem}
\newtheorem{ctheorem}{Corrected Theorem}
\newtheorem{corollary}{Corollary}
\newtheorem{proposition}{Proposition}
\newtheorem{example}{Example}
\newtheorem{algorithm}{\underline{Algorithm}}[section]
\newcommand{\mat}[2]{\ensuremath{
\left( \begin{array}{c} #1 \\ #2 \end{array} \right)}}
\newcommand{\eq}[1]{(\ref{#1})}
\newcommand{\one}[1]{\ensuremath{\mathbf{1}_{#1}}}
\newcommand{\am}{\mbox{argmin}}
\newcommand{\dmin}{d_{\mbox{min}}}
\newcommand{\be}{\begin{equation}}
\newcommand{\ee}{\end{equation}}
\newcommand{\eps}{\varepsilon}
\newcommand{\imipi}{\int_{-\infty}^{\infty}}
\newcommand{\mug}{\stackrel{\triangle}{=}}
\renewcommand{\thesubsection}{\Alph{subsection}}
\def \bfpi  {\bm{\pi}}
\def \bflambda  {\bm{\lambda}}

\newcommand{\mcA}{\mathcal{A}}
\newcommand{\mcB}{\mathcal{B}}
\newcommand{\mcS}{\mathcal{S}}
\newcommand{\mcN}{\mathcal{N}}
\newcommand{\integers}{\mathbb{Z}}
\newcommand{\naturals}{\mathbb{N}}
\newcommand{\Vmn}{V_{m\times n}}
\newcommand{\zmn}{z_{m\times n}}
\newcommand{\mtn}{m\times n}
\newcommand{\mtnn}{m{\times} n}  
\newcommand{\mbPi}{\mathbf{\Pi}}
\newcommand{\mbm}{\mathbf{m}}
\newcommand{\xhmn}{\hat{x}_{\mtnn}}
\newcommand{\xmn}{x_{\mtnn}}
\newcommand{\Xmnuniv}{\hat{X}^{\mtnn}_{{\scriptsize{\sf univ}}}}
\newcommand{\mexp}[2]{#1{\cdot} 10^{#2}}
\renewcommand{\thesubsection}{\thesection-\Alph{subsection}}

\section{Introduction}

The problem of estimating a discrete-time, finite-alphabet source
signal $\{X_t\}_{t\in T}$ from the entire observation of a noisy
signal $\{Z_t\}_{t\in T}$, which has been corrupted by a known
discrete memoryless channel (DMC), has been thoroughly studied
recently in \cite{DUDE}. It has been shown that even though the
source distribution is unknown, an algorithm called DUDE can
universally achieve the asymptotically optimal performance. This
result has been extended in various directions such as the case of
channel uncertainty \cite{gemelos04}, the case where the channel has
memory \cite{zhang05}, the case of non-discrete noisy signal
components \cite{DemboWeissmancontoutputdude2003}, and the case
where the reconstruction is required to depend causally on the noisy
signal \cite{sequentialdude04}\cite{filteringbyprediction}. In this
paper, we revisit the last case, taking a different approach from
\cite{sequentialdude04}\cite{filteringbyprediction}.

The case where we estimate $X_t$ causally based on observation of
the noisy signal $Z^t=(Z_1,\cdots,Z_t)$, is referred to as
\textit{filtering}. The filter can be either deterministic or
randomized (a concept that will be explained in detail later). In
this paper, we will only focus on the \textit{stochastic setting},
where we assume $\{X_t\}$ is a stationary and ergodic stochastic
process. With the stochastic setting assumption, and under the same
performance criterion of \cite{DUDE}, i.e., minimizing the expected
normalized cumulative loss, knowledge of the conditional
distribution of $X_t$ given $Z^t$ at each time $t$ is required to
achieve the optimal performance. Also, by the same argument as in
\cite[Section III]{DUDE}, this conditional distribution can be
obtained by the conditional distribution of $Z_t$ given $Z^{t-1}$
when the invertible DMC is known. (We call a channel is invertible
if its transition probability matrix is invertible.)

However, for the \textit{universal} filtering setting, where the
probability distribution of the source is unknown, the conditional
distribution of $Z_t$ given $Z^{t-1}$ is also not known and need be
learned from the observed noisy signal. Therefore, if we can learn
this conditional distribution accurately as the observation length
increases, we can hope to build the universal filtering scheme that
achieves the asymptotically optimal performance from the estimated
conditional distribution. To pursue this goal,
\cite{sequentialdude04}\cite{filteringbyprediction} adopt the
universal prediction\cite{MerhavGutmanFeder92} approach. That is,
they first get an estimate of the conditional distribution of $Z_t$
given $Z^{t-1}$ by employing a universal predictor for the observed
noisy signal, and then by inverting the known DMC, obtain an
estimate of the conditional distribution of $X_t$ given $Z^t$.

Unlike the approach of
\cite{sequentialdude04}\cite{filteringbyprediction}, in this work,
we turn our attention to the rich theory of hidden Markov process
(HMP) models to directly obtain a different kind of estimate of the
conditional distribution of $X_t$ given $Z^t$, without going through
the channel inversion stage.

Generally, HMPs are defined as a family of stochastic processes that
are outputs of a memoryless channel whose inputs are finite state
Markov chains. As can be seen in \cite{EphraimMerhav2002}, these HMP
models arise in many areas, such as information theory,
communications, statistics, learning, and speech recognition. Among
these applications of HMPs, there are many situations where the
state of the underlying Markov chain need be estimated based on the
observed hidden Markov process. If the exact parameters of the HMP,
namely, the state transition probability of the Markov chain and the
channel transition density, as well as the order of the Markov chain
are known, then this problem can be easily solved via well-known
forward-backward recursions which were discovered by
\cite{ChangHancock66} and \cite{BPSW70}. Especially, when we are
estimating the state based on the causal observation of the HMP, we
only need the forward recursion formula. In addition, much work has
been done for the state estimation, where the order is known, but
the parameters of the HMP are unknown. In this case, the parameters
are first estimated via maximum likelihood (ML) estimation or the EM
algorithm, then the state is estimated by using the estimated
parameters in the recursion formula. A detailed explanation of this
approach and the property of the ML parameter estimation can be
found in \cite{BPSW70}\cite{BRR98}\cite{Leroux92}\cite{Finesso90}.
Furthermore, this was extended to the case where the order of the
Markov chain is also not known, but the upper bound on the order is
known. In this case, the order estimation is first performed before
the parameter and state estimation, and the above process is
repeated. The references for the order estimation are given in
\cite{Kieffer93}\cite{LN94}\cite{Ryden95}. There also has been work
for the case where even the knowledge of the upper bound on the
order of the Markov chain is not
required\cite{Finesso90}\cite{ZivMerhav92}.

From these rich theories for the state, parameter, and order
estimation of HMPs, we can see that it is possible to build a
universal filtering scheme if the source distribution is known to be
a finite state Markov chain. That is, since the channel is
memoryless and fixed in our setting, if our source $\{X_t\}$ is a
finite state Markov chain, then obviously, $\{Z_t\}$ is an HMP, and
we can first estimate the order of the Markov chain, then estimate
the parameter, and finally perform forward recursion to learn the
conditional distribution of $X_t$ given $Z^t$. From the consistency
results of order estimation and parameter estimation, this
conditional distribution will be an accurate estimate of the true
one, and we can use it to build the universal filtering scheme.

Now, in our work, we extend this approach to the case where our
source $\{X_t\}$ is a general stationary and ergodic process (with
some benign conditions), which need not be a Markov source at all,
and show that we can still build a universal filtering scheme that
achieves asymptotically optimal performance. The skeleton of our
scheme is the following: We first ``model'' our source as a finite
state Markov chain with a certain order, or equivalently, model the
noisy observed signal $\{Z_t\}$ as an HMP in a certain class. Then,
we estimate the parameters of the HMP that ``approximates'' the
noisy signal best in that class. We will show that from the
consistency result about the ML parameter estimation for the
mismatched model \cite{Finesso90}, these estimated parameters will
give an accurate estimation of the conditional distribution of $X_t$
given $Z^t$, as the observation length increases and the HMP class
gets richer. Then, this result will guarantee that our universal
filter using this conditional distribution will attain the
asymptotically optimal performance. In practice, this approach has
been heuristically employed in many applications for nonlinear
filtering without theoretical justification. Therefore, this work
shows the first theoretical proof on the justification of the
HMP-based universal filtering scheme.

The remainder of the paper is organized as follows. Section
\ref{sec: notation} introduces some notation and preliminaries that
are needed for setting up the problem. In Section \ref{sec:
universal filtering}, the universal filtering problem is defined
explicitly. In Section \ref{sec: universal filtering via HMM}, our
universal filtering scheme is devised, the main theorem is stated,
and proved. Section \ref{sec: extension} extends our approach to the
case where the channel has memory. Section \ref{sec: conclusion}
concludes the paper and lists some related future directions.
Detailed technical proofs that are needed in the course of proving
our main results are given in the Appendix.

\section{Notation and preliminaries} \label{sec: notation}

\subsection{General notation} \label{sec: Problem Setting}

We assume that the clean, noisy and reconstruction signal components
take their values in the same finite $M$-ary alphabet
$\mathcal{A}=\{0,\cdots,M-1\}$. The simplex of $M$-dimensional
column probability vectors will be denoted as $\mathcal{M}$.

The DMC is known to the filter and is denoted by its transition
probability matrix $\mathbf{\Pi}=\{\Pi(i,j)\}_{i,j\in\mathcal{A}}$.
Here, $\Pi(i,j)$ denotes the probability of channel output symbol
$j$ when the input is $i$. We assume $\Pi(i,j)>0$ $\forall i,j$, and
let $\Pi_{\min}=\min_{i,j}\Pi(i,j)$. We assume this channel matrix
is invertible and denote the inverse as $\mathbf{\Pi}^{-1}$. Let
$\Pi_i^{-1}$ denote the $i$-th column of $\mathbf{\Pi}^{-1}$. We
also assume a given loss function (fidelity criterion)
$\Lambda:\mathcal{A}^2\rightarrow[0,\infty)$, represented by the
loss matrix $\mathbf{\Lambda}=\{\Lambda(i,j)\}_{i,j\in\mathcal{A}}$,
where $\Lambda(i,j)$ denotes the loss incurred when estimating the
symbol $i$ with the symbol $j$. The maximum single-letter loss will
be denoted by $\Lambda_{\max}=\max_{i,j\in\mathcal{A}}\Lambda(i,j)$,
and $\lambda_j$ will denote the $j$-th column of $\mathbf{\Lambda}$.

As in \cite{DUDE}, we define the extended Bayes response associated
with the loss matrix $\mathbf{\Lambda}$ to any column vector
$\mathbf{V}\in\mathbb{R}^{M}$ as
$$
B(\mathbf{V})=\arg\min_{\hat{x}\in\mathcal{A}}\lambda_{\hat{x}}^{T}\mathbf{V},
$$
where $\arg\min_{\hat{x}\in\mathcal{A}}$ denotes the minimizing
argument, resolving ties by taking the letter in the alphabet with
the lowest index.

We let $P$ denote the true joint probability law of the clean and
noisy signal, and $E(\cdot)$ denote expectation with respect to $P$.
Also, every almost sure convergence is with respect to $P$. If we
need to refer to the probability law of clean or noisy signal
induced by $P$, we denote $P_X$ and $P_Z$, respectively. If $P$ is
written in a bold face, $\mathbf{P}$, with a subscript, it stands
for a simplex vector in $\mathcal{M}$ for the corresponding
distribution of the subscript. For example, $\mathbf{P}_{X_t|z^t}$
is a column $M$-vector whose $i$-th component is $P(X_t=i|Z^t=z^t)$.

When we have some other probability law denoted as $Q$, and want to
measure its difference from $P$, a natural choice of such a measure
is the relative entropy rate. First, denote the $n$-th order
relative entropy between $P$ and $Q$ as
\begin{align*}
D_{n}(P||Q) =&
\sum_{z^{n}}P(z^{n})\log\frac{P(z^{n})}{Q(z^{n})}=E\Big(\log\frac{P(Z^n)}{Q(Z^n)}\Big).
\end{align*}
Then, the relative entropy rate (also known as Kullback-Leibler
divergence rate) is defined as
$$
\mathbf{D}(P\|Q)\triangleq\lim_{n\rightarrow\infty}\frac{1}{n}D_{n}(P||Q)
$$
if the limit exists. When $Q$ is a probability law in a certain
class of HMPs, this limit always exists and the relative entropy
rate is well defined. A more detailed discussion about this limit
will be given in Lemma \ref{lemma for KL divergence}.
This relative entropy rate
will play a central role in analyzing our universal filtering
scheme.

\subsection{Hidden Markov Processes (HMPs)}
\subsubsection{Definition}\label{Hidden Markov Processes}

As stated in the Introduction, the HMPs are generally defined as a
family of stochastic processes that are outputs of a memoryless
channel whose inputs are finite
state Markov chains. 
Throughout the paper, we will only consider the case in which the
alphabet of HMP, $\mathcal{Z}$, and underlying Markov chain,
$\mathcal{X}$, are finite and equal, i.e.,
$\mathcal{Z}=\mathcal{X}=\mathcal{A}$, and the channel is DMC and
invertible.


There are three parameters that determine the probability laws of
HMP: $\pi$, the initial distribution of finite state Markov chain;
$A$, the probability transition matrix of finite state Markov chain,
and $B$, the probability transition matrix of DMC. The triplet
$\{\pi,A,B\}$ is referred to as the parameter of HMP. Let $\Theta$
be a set of all $\theta$'s where $\theta :=
\{\pi_{\theta},A_{\theta},B_{\theta}\}$. For each $\theta$, we can
calculate the likelihood function
\begin{displaymath}
Q_{\theta}(z^{n})=\pi_{\theta}\prod_{t=1}^{n}(\hat{\textbf{B}}_{\theta,t}A_{\theta})\textbf{1},
\end{displaymath}
where $\hat{\textbf{B}}_{\theta,t}$ is $M \times M$ diagonal matrix
whose $(j,j)$-th entry is the $(j,z_{t})$-th entry of $B_{\theta}$,
and $\textbf{1}$ is the $M\times1$ vector with all entries equal to
$1$.

\indent Now, let $\Theta_{k}\subset\Theta$ be a set of $\theta$'s,
such that the order of underlying Markov chain of HMP is $k$.
Furthermore, for some $\delta>0$, define
$\Theta_k^{\delta}\subset\Theta_{k}$ as the set of
$\theta\in\Theta_k$ satisfying:
\begin{itemize}
\item $a_{ij,\theta}\geq\delta$, if the first $k-1$ components of the $k$-tuple state $j$ are equal to the last $k-1$ components of $k$-tuple state $i$
\item $a_{ij,\theta}=0$, otherwise
\item $b_{ij,\theta}=\mathbf{\Pi}(i,j)$, for $\forall i,j$,
\end{itemize}
where $a_{ij,\theta}$ is $(i,j)$-th entry of $A_{\theta}$, and
$b_{ij,\theta}$ is $(i,j)$-th entry of $B_{\theta}$. In particular,
if $\theta\in\Theta_{k}^{\delta}$ then: 1) the stochastic matrix
$A_{\theta}$ is irreducible and aperiodic; thus, if the Markov chain
is stationary, $\pi_{\theta}$ is the stationary distribution of the
Markov chain, and is uniquely determined from $A_{\theta}$, 2)
$B_{\theta}=\mathbf{\Pi}$ $\forall \theta$, and, therefore, $\theta$
is completely specified by $A_{\theta}$. For notational brevity, we
omit the subscript $\theta$ and denote the probability law
$Q\in\Theta_{k}^{\delta}$, if $Q=Q_{\theta}$, and
$\theta\in\Theta_{k}^{\delta}$.

\subsubsection{Maximum likelihood (ML) estimation}

Generally, suppose a probability law $Q$ is in a certain class
$\Omega$. Then, the $n$-th order maximum likelihood (ML) estimator
in $\Omega$ for the observed sequence $z^n$, is defined as
$$
\hat{Q}[z^n]=\arg\max_{Q\in\Omega} Q(z^{n}),
$$
resolving ties arbitrarily. Now, if $Q\in\Theta_{k}^{\delta}$, then
there is an algorithm called expectation-maximization(EM) [4] that
iteratively updates the parameter estimates to maximize the
likelihood. Thus, when $Q$ is in the class of probability laws of a
HMP, the maximum likelihood estimate can be efficiently
attained.\footnote{We neglect issues of convergence of the EM
algorithm and assume that the ML estimation is performed perfectly.}
We denote the ML estimator in $\Theta_{k}^{\delta}$ based on $z^n$
by
$$
\hat{Q}_{k,\delta}[z^n]=\arg\max_{Q\in\Theta_{k}^{\delta}} Q(z^{n}).
$$
Obviously, when the n-tuple $Z^n$ is random,
$\hat{Q}_{k,\delta}[Z^n]$ is also a random probability law that is a
function of $Z^n$.

\subsubsection{Consistency of ML estimator}\label{consistency}

When $P_Z\in\Theta_k^{\delta}$, an ML estimator
$\hat{Q}_{k,\delta}[Z^n]$ is said to be \emph{strongly consistent}
if
$$
\lim_{n\rightarrow\infty}\hat{Q}_{k,\delta}[Z^n]=P_Z\qquad\textrm{\emph{a.s.}}
$$
The strong consistency of the ML estimator $\hat{Q}_{k,\delta}[Z^n]$
of the parameter of a finite-alphabet stationary ergodic HMP was
proved in \cite{BaumPetrie66}. For the case of a general stationary
ergodic HMP, the strong consistency was proved in \cite{Leroux92}.

We also have a sense of strong consistency for the case where $P_Z$
is a general stationary and ergodic process. By
the similar argument as in \cite[Theorem 2.2.1]{Finesso90}, we have
the consistency in the sense that if the observed noisy signal is
not necessarily a HMP, and we still perform the ML estimation in
$\Theta_k^{\delta}$, then we get
\begin{align}
\lim_{n\rightarrow\infty}\hat{Q}_{k,\delta}[Z^n]\in
\mathcal{N}\qquad\textrm{\emph{a.s.}},\label{second consistency}
\end{align}
where
$\mathcal{N}\triangleq\{Q\in\Theta_k^{\delta}:\mathbf{D}(P\|Q)=\min_{Q^{'}\in\Theta_k^{\delta}}\mathbf{D}(P\|Q^{'})\}$.\footnote{Just
as in \cite[Theorem 2.2.1]{Finesso90}, the notion of a.s. set
convergence is used. For any subset $\mathcal{E}\in\Theta$, define
$\mathcal{E}_{\epsilon}\triangleq\{Q\in\Theta:d(Q,\mathcal{E})<\epsilon\}$,
where $d$ is the Euclidean distance. Then,
$\lim_{n\rightarrow\infty}\hat{Q}[Z^n]\in\mathcal{E}$ a.s. if
$\forall\epsilon>0, \exists N(\epsilon,\omega)$ such that $\forall
n\geq N(\epsilon,\omega), \hat{Q}[Z^n]\in\mathcal{E_{\epsilon}}$}
This second consistency result is the key result that we will use in
devising and analyzing our universal filtering scheme.

\section{The universal filtering problem}\label{sec: universal
filtering}

As mentioned in the Introduction, we will assume a stochastic
setting, that is, the underlying clean signal is an output of some
stationary and ergodic process whose probability law is $P_X$. From
$P_X$ and $\mathbf{\Pi}$, we can get the true joint probability law
$P$ and corresponding probability law of noisy observed signal,
$P_Z$. That is,
\begin{eqnarray*}
P(X^n=x^n,Z^n=z^n)&=&P_X(X^n=x^n)\prod_{t=1}^{n}\Pi(x_i,z_i),\quad\textrm{
and}\\
P_Z(Z^n=z^n)&=&\sum_{x^n}P(X^n=x^n,Z^n=z^n).
\end{eqnarray*}

A \emph{filter} is a sequence of probability distributions
$\hat{\textbf{X}}= \{\hat{X}_{t}\}$, where
$\hat{X}_{t}:\mathcal{A}^{t}\rightarrow\mathcal{M}$. The
interpretation is that, upon observing $z^{t}$, the reconstruction
for the underlying, unobserved $x_{t}$ is represented by the symbol
$\hat{x}$ with probability $\hat{X}_{t}(z^{t})[\hat{x}]$. A filter
is called \emph{deterministic} if $\hat{X}_{t}(z^{t})$ is a unit
vector in $\mathbb{R^M}$ for all $t$ and $z^t$, and
\emph{randomized} if $\hat{X}_{t}(z^{t})$ can be a simplex vector in
$\mathcal{M}$ other than a unit vector for some $t$ and $z^t$. The
\textit{normalized cumulative loss} of the scheme $\hat{\textbf{X}}$
on the individual pair $(x^{n},z^{n})$ is defined by
\begin{displaymath}
L_{\hat{\textbf{X}}}(x^{n},z^{n})=\frac{1}{n}\sum_{t=1}^{n}\ell(x_{t},\hat{X}_{t}(z^{t})),
\end{displaymath}
where $
\ell(x_{t},\hat{X}_{t}(z^{t}))=\sum_{\hat{x}\in\mathcal{\hat{X}}}\Lambda(x_{t},\hat{x})\hat{X}_{t}(z^{t})[\hat{x}].
$ Then, the goal of a filter is to minimize the \textit{expected
normalized cumulative loss}
$E\Big(L_{\hat{\textbf{X}}}(X^{n},Z^{n})\Big)$.

The optimal performance of the $n$-th order filter is defined as
$$
\phi_{n}(P_X,\mathbf{\Pi})=\min_{\hat{\mathbf{X}}\in\mathcal{F}}E\Big(L_{\hat{\mathbf{X}}}(X^{n},Z^{n})\Big),
$$
where $\mathcal{F}$ denotes the class of all filters. Sub-additivity
arguments similar to those in \cite{DUDE} imply
\begin{displaymath}
\lim_{n\rightarrow\infty}\phi_{n}(P_X,\mathbf{\Pi})=\inf_{n\geq1}\phi_{n}(P_X,\mathbf{\Pi})\triangleq\mathbf{\Phi}(P_X,\mathbf{\Pi}).
\end{displaymath}
By definition, $\mathbf{\Phi}(P_X,\mathbf{\Pi})$ is the
(distribution-dependent) optimal asymptotic filtering performance
attainable when the clean signal is generated by the law $P_X$ and
corrupted by $\mathbf{\Pi}$. This $\mathbf{\Phi}(P_X,\mathbf{\Pi})$
can be achieved by the optimal filter
$\hat{\mathbf{X}}_{P}=\{\hat{X}_{P,t}\}$ where
$$
\hat{X}_{P,t}(z^{t})[\hat{x}]=Pr(B(\mathbf{P}_{X_t|z^t})=\hat{x}).
$$
\noindent For brevity of notation, we denote
$\hat{X}_{P}(z^{t})=\hat{X}_{P,t}(z^{t})$. Note that this is a
\textit{deterministic} filter, i.e., for a given $z^t$, the filter
is a unit vector in $\mathbb{R}^M$ for all $t$. We can easily see
that this filter is optimal since it minimizes
$E(\ell(X_t,\hat{X}(Z^t))$ for all $t$, and thus, it minimizes
$E\Big(L_{\hat{\textbf{X}}}(X^{n},Z^{n})\Big)$ for all $n$.

As can be seen, $\hat{X}_{P}(z^{t})$ needs the exact knowledge of
$\mathbf{P}_{X_t|z^t}$, and thus, is dependent on the distribution
of the underlying clean signal. The \emph{universal filtering
problem} is to construct (possibly a sequence of) filter(s),
$\hat{\mathbf{X}}_{univ}$, that is independent of the distribution
of underlying clean signal, $P_X$, and yet asymptotically achieving
$\mathbf{\Phi}(P_X,\mathbf{\Pi})$. We describe our sequence of
universal filters in the next section.

%
%
%

\section{Universal filtering based on hidden Markov
modeling}\label{sec: universal filtering via HMM}
\subsection{Description of the filter}\label{sec: description of the
filter} Before describing our sequence of universal filters, we make
the following assumption on the source.
\begin{assumption}
There exists a sequence of positive reals $\{\delta_k\}$, such that
$\delta_k\downarrow0$ as $k\rightarrow\infty$, and $P_X$ satisfies
\begin{align}
P_X(X_0|X_{-k}^{-1})\geq\delta_k \quad\textrm{a.s.}\quad \forall
k\in\mathbb{N}.\label{assumption of source}
\end{align}
\end{assumption}


For any probability law $Q$, we construct a \textit{randomized}
filer as follows: For $\epsilon>0$, denote $L_2$ $\epsilon$-ball in
$\mathbb{R}^M$ as $
B_{\epsilon}=\{\mathbf{V}\in\mathbb{R}^M:\|\mathbf{V}\|_2\leq\epsilon\}$.
Then, we define a filter for fixed $\epsilon$ as
\begin{align}
\hat{X}_{Q,t}^{\epsilon}(z^{t})[\hat{x}]=Pr(B(\mathbf{Q}_{X_t|z^t}+\mathbf{U})=\hat{x}),\label{randomized
filter}
\end{align}
%
where $\mathbf{U}\in\mathbb{R}^M$ is a random vector, uniformly
distributed in $B_{\epsilon}$. For brevity of notation, we denote
$\hat{X}_{Q}^{\epsilon}(z^{t})=\hat{X}_{Q,t}^{\epsilon}(z^{t})$.
This filter is randomized since depending on $Q$ and $z^t$,
$\hat{X}_{Q}^{\epsilon}(z^{t})$ can be a probability simplex vector
in $\mathcal{M}$ that is not a unit vector. The reason we needed
this randomization will be explained in proving Lemma \ref{lemma for
concentration}.

To devise our filter, let's first consider an increasing sequence of
positive integers, $\{m_i\}_{i\geq1}$, that satisfies following
conditions:
\begin{align}
\lim_{i\rightarrow\infty}\frac{m_{i-1}}{m_i}=0,\quad\lim_{i\rightarrow\infty}m_i=\infty.\label{condition
on sequence}
\end{align}
Now, define
$$
i(t)\triangleq\max\{i:m_i\leq t\}.
$$
Then, given that our source distribution satisfies \eq{assumption of
source}, and for fixed $k$, define a random probability law
\begin{align}
Q_{k}^{t}\triangleq&
\hat{Q}_{k,\delta_k}[Z^{m_{i(t)}}]=\arg\max_{Q\in\Theta_{k}^{\delta_k}}
Q(Z^{m_{i(t)}}).\label{random Q}
\end{align}
That is, $Q_k^t$ is the ML estimator in $\Theta_{k}^{\delta_k}$
based on $Z^{m_{i(t)}}$. As discussed in Section \ref{Hidden Markov
Processes}, we only need to estimate the state transition
probabilities of the underlying Markov chain to obtain this ML
estimator, and this can be efficiently done by the
Expectation-Maximization (EM) algorithm. Once we get $Q_k^t$, we can
then calculate $\mathbf{Q}^t_{k X_t|z^t}$ using the
forward-recursion formula which is described in detail in [4]. Note
that we get this conditional distribution directly, not by first
estimating the output distribution, and then inverting the channel,
as was done in
\cite{sequentialdude04}\cite{filteringbyprediction}\cite{DUDE}.


Finally, we take as our sequence of universal filtering schemes,
indexed by $k$ and $\epsilon$,
$$
\hat{\mathbf{X}}_{univ,k}^{\epsilon}=\{\hat{X}_{Q_k^t,t}^{\epsilon}\}.
$$
The following theorem states the main result of this paper.

\begin{theorem}\label{main theorem}
Let $\mathbf{X}^{\infty}\in \mathcal{A}^{\infty}$ be a stationary,
ergodic process emitted by the source $P_X$ which satisfies
Assumption 1. Let $\mathbf{Z}^{\infty}\in \mathcal{A}^{\infty}$ be
the output of the DMC, $\mathbf{\Pi}$, whose input is
$\mathbf{X}^{\infty}$. Then:
\begin{itemize}
\item[(a)]
$\lim_{\epsilon\rightarrow0}\lim_{k\rightarrow\infty}\limsup_{n\rightarrow\infty}L_{\hat{\mathbf{X}}_{univ,k}^{\epsilon}}(X^{n},Z^{n})\leq\mathbf{\Phi}(P_X,\Pi)$
\quad a.s.
\item[(b)]
$\lim_{\epsilon\rightarrow0}\lim_{k\rightarrow\infty}\limsup_{n\rightarrow\infty}E\Big(L_{\hat{\mathbf{X}}_{univ,k}^{\epsilon}}(X^{n},Z^{n})\Big)=\mathbf{\Phi}(P_X,\Pi)$
\end{itemize}
\end{theorem}

\subsection{Intuition behind the scheme and proof
sketch}\label{sec: intuition}

The intuition behind our scheme parallels that of the universal
compression and universal prediction problems in the stochastic
setting. In the $n$-th order problem of both cases
\cite{CoverThomas91}\cite{Merhav1998}, the excess expected codeword
length per symbol and the excess expected normalized cumulative loss
incurred by using the wrong probability law $Q$ in place of the true
probability law $P$ could be upper bounded by the normalized $n$-th
order relative entropy $\frac{1}{n}D_n(P\|Q)$. Then, to achieve the
asymptotically optimum performance, the compressor and the predictor
try to find and use some data-dependent $Q$ that makes
$\frac{1}{n}D_n(P\|Q)\rightarrow0$ as $n\rightarrow\infty$, that is,
makes $\mathbf{D}(P\|Q)$ zero.

We follow the same intuition in our universal filtering problem. For
fixed $k$ and $\epsilon$, our scheme, as can be seen from \eq{random
Q}, divides the noisy observed signal into sub-blocks of length
$(m_i-m_{i-1})$. Since $\frac{m_{i-1}}{m_i}$ tends to zero as
$i\rightarrow\infty$, the length of each sub-block grows faster than
exponential. Now, to filter each sub-block, it plugs the ML
estimator in $\Theta_k^{\delta_k}$ obtained from the entire
observation of noisy signal up to the previous sub-block. From
\eq{second consistency}, we know that as the observation length $n$
increases, this ML estimator will converge to the parameter that
minimizes the relative entropy rate between the true output
probability law $P_Z$. Then, to show that this scheme achieves the
asymptotically optimum performance, we bound the excess expected
normalized cumulative loss with this relative entropy rate, and show
that the bound goes to zero as the HMP parameter set becomes richer,
that is, $k$ increases.

%
%
%
%

To be more specific, we briefly sketch the proof of our main
theorem. Part (b) of Theorem \ref{main theorem} states that our
scheme is asymptotically optimal. We can easily see that this
follows directly from Part (a) and Reverse Fatou's Lemma. Therefore,
proving Part (a) is the key in proving the theorem. Part (a) states
that in the limit, the normalized cumulative loss of our scheme, for
almost every realization, is less than or equal to the
asymptotically optimum performance.

To prove Part (a), we first fix $k$ and $\epsilon$, and get the
following inequality
\begin{align}
\limsup_{n\rightarrow\infty}\Big(L_{\hat{\mathbf{X}}_{univ,k}^{\epsilon}}(X^n,Z^n)-\phi_n(P_X,\mathbf{\Pi})\Big)\leq
F\Big(\limsup_{t\rightarrow\infty}\mathbf{D}(P_Z\|Q_k^t),\epsilon\Big)\qquad\textrm{\emph{a.s.}},\label{proof
sketch}
\end{align}
where $F(x,y)$ is some function such that $F(x,y)\rightarrow0$ as
$x\downarrow0$, and then $y\downarrow0$.\footnote{Note that $Q_k^t$
in $\mathbf{D}(P_Z\|Q_k^t)$ is a function of $Z^{m_{i(t)}}$, and
thus, is random. A more formal definition of relative entropy rate
between true and the random probability law like this case will be
given after Lemma \ref{lemma for difference of expected loss}.}
There are two keys in getting this inequality. The first one is to
show the concentration of
$L_{\hat{\mathbf{X}}_{univ,k}^{\epsilon}}(X^n,Z^n)$ to its
expectation which will be shown in Lemma \ref{lemma for
concentration} and Corollary \ref{univ concentration}. The second is
to get the explicit upper bound function $F(x,y)$ which will be
based on Lemma \ref{lemma for difference of expected loss}. Once
establishing this inequality, we show that
\begin{align}
\lim_{k\rightarrow\infty}\limsup_{t\rightarrow\infty}\mathbf{D}(P_Z\|Q_k^t)=0\qquad\textrm{\emph{a.s.}},\label{upper
bound}
\end{align}
from Lemma \ref{lemma for two inequalities} and then send
$\epsilon\downarrow0$ to get Part (a). Keeping this proof sketch in
mind, let us move on to the detailed proof in the next section.

\subsection{Proof of the theorem}
Before proving the theorem, we introduce several lemmas as building
blocks. Lemma \ref{lemma for uniform convergence} and Lemma
\ref{lemma for KL divergence} below give some general results for
the HMPs that we are considering. Our lemmas are similar to
\cite[Lemma 2.3.4]{Finesso90} and \cite[Theorem 2.3.3]{Finesso90}.
The latter assumed that all the parameters are lower bounded by
$\delta>0$, whereas in $\Theta_k^{\delta}$, some parameters can be
zero. We take this into account in proving Lemma \ref{lemma for
uniform convergence} and Lemma \ref{lemma for KL divergence}. Lemma
\ref{lemma for concentration} shows the uniform concentration
property of the normalized cumulative loss on $\Theta_k^{\delta}$,
which is an important property that we need to prove the main
theorem. Lemma \ref{lemma for difference of expected loss} provides
a key step to get the upper bound described in \eq{proof sketch},
and Lemma \ref{lemma for two inequalities}, which needs three
additional definitions, enables to show \eq{upper bound}. After
building up the lemmas, we give the proof of the main theorem, which
is merely an application of the lemmas.

\begin{lem}\label{lemma for uniform convergence}
Suppose $Q\in\Theta_k^{\delta}$ and fix $\delta>0$. Then, $\forall
\omega$, $Q(Z_{0}|Z_{-t}^{-1})$ converges to a limit
$Q(Z_{0}|Z_{-\infty}^{-1})$ uniformly on $\Theta_{k}^{\delta}$.
\end{lem}

\emph{Proof:} To prove this lemma, we need three more lemmas in
Appendix 1, which are variations on those found in
\cite{BaumPetrie66}. Let's denote $f_t:=Q(Z_0|Z_{-t}^{-1})$, and
$f_0=0$. Then, the sequence $\{f_t\}$ uniformly converges on
$\Theta_{k}^{\delta}$, if following $k$ subsequences,
$$
\{f_{jk+l}, j=0,1,2,\cdots,\},\qquad 0\leq l\leq k-1,
$$
uniformly converge on $\Theta_{k}^{\delta}$, and have the same
limit.

First, the uniform convergence of each subsequence $\{f_{jk+l}\}$
can be shown by showing the series
$\sum_{j=0}^{t}(f_{(j+1)k+l}-f_{jk+l})$ converges absolutely. From
Lemma \ref{third lemma for uniform convergence} in Appendix 1,
setting $m=k$,
\begin{align*}
 &\sum_{j=0}^{t}|f_{(j+1)k+l}-f_{jk+l}|\nonumber\\
=&\sum_{x_0}Q(Z_{0}|x_{0})\sum_{j=1}^{t}|Q(x_{0}|Z_{-(j+1)k-l}^{-1})-Q(x_{0}|Z_{-jk-l}^{-1})|\nonumber\\
\leq&M\sum_{j=1}^{t}(\rho_{\delta,k,k})^{j+1}.\nonumber
\end{align*}
Since $\rho_{\delta,k,k}<1, M<\infty$ and $\rho_{\delta,k,k}$ does
not depend on $Q$, $\omega$, and $l$, we conclude that all $k$
subsequences converge uniformly on $\Theta_{k}^{\delta}$.

Now, to show that the $k$ subsequences have the same limit,
construct another subsequence, $\{f_{j(k+1)+1}, j=0,1,2,\cdots,\}$.
Since this subsequence contains infinitely many terms from all $k$
subsequences, if this subsequence converges uniformly on
$\Theta_{k}^{\delta}$, we can conclude that the $k$ subsequences
have the same limit. The derivation of the uniform convergence of
this subsequence is the same as that described above, but setting
$m=k+1$ in Lemma 8. Therefore, the original sequence $\{f_t\}$
converges to its limit uniformly on $\Theta_{k}^{\delta}$.\quad
$\blacksquare$

The remarkable fact of this lemma is that the convergence is not
only uniform on $\Theta_k^{\delta}$, but also in $\omega$. That is,
the convergence holds uniformly on every realization of
$z_{-\infty}^{0}$.

\begin{lem}\label{lemma for KL divergence}
For the distribution of the observed noisy process $\{Z_t\}$, $P_Z$,
and every $Q\in\Theta_k^{\delta}$,
$$
\mathbf{D}(P_Z\|Q)\triangleq\lim_{n\rightarrow\infty}\frac{1}{n}D_{n}(P_Z\|Q)=E\Big(\log\frac{P_Z(Z_{0}|Z_{-\infty}^{-1})}{Q(Z_{0}|Z_{-\infty}^{-1})}\Big).
$$
Moreover,
$$
\lim_{n\rightarrow\infty}\frac{1}{n}\log\frac{P_Z(Z^n)}{Q(Z^n)}=\mathbf{D}(P_Z\|Q)\qquad\textrm{
 \emph{a.s.} \quad uniformly on $\Theta_k^{\delta}$.}
$$
\end{lem}

\emph{Proof:} This lemma consists of three parts. The first part is
to show the existence of the first limit in the lemma so that the
definition of $\mathbf{D}(P_Z\|Q)$ is valid. The second part is to
show that the value of the limit is indeed
$E\Big(\log\frac{P_Z(Z_{0}|Z_{-\infty}^{-1})}{Q(Z_{0}|Z_{-\infty}^{-1})}\Big)$.
Finally, the last part is to show the uniform convergence of
normalized log-likelihood ratio to the relative entropy rate. The
first two parts and the pointwise convergence of the third part is a
generalization of the Shannon-McMillan-Breiman theorem. The proof of
these parts is identical to those in \cite[Theorem 2.3.3]{Finesso90}
even for the case where some parameters in $\Theta_k^{\delta}$ can
be zero.

The uniform convergence in the third part of the lemma is crucial in
that it enables to obtain the second consistency result \eq{second
consistency} as in \cite[Theorem 2.2.1]{Finesso90}. We take into
account our parameter set, and repeat the argument of \cite[Lemma
2.4.1]{Finesso90}. To show the uniform convergence, we need to show
$$\lim_{n\rightarrow\infty}\frac{1}{n}\log Q(Z^n)=E\Big(\log
Q(Z_0|Z_{-\infty}^{0})\Big)\qquad \textrm{ \emph{a.s.} \quad
uniformly on $\Theta_k^{\delta}$}$$  Since the pointwise convergence
can be shown and the parameter set $\Theta_{k}^{\delta}$ is compact,
it is enough to show that $\frac{1}{n}\log Q(Z^n)$ is an
equicontinuous sequence by Ascoli's Theorem. That is, we need to
show for $\forall\epsilon>0$, $\exists\delta(\epsilon)>0$ such that
$$
\forall n, \left|\frac{1}{n}\log Q(Z^n)-\frac{1}{n}\log
Q^{'}(Z^n)\right|\leq\epsilon, \textrm{  if}\quad
\|Q-Q^{'}\|_{1}<\delta(\epsilon),
$$
where $\|Q-Q^{'}\|_{1}\triangleq\sum_{i,j}|a_{ij}-a^{'}_{ij}|$ is
defined to be the $L_1$ distance between the two parameters defining
$Q$ and $Q^{'}$. This equicontinuity can be proved by observing that
a process $\{S_{t}=(X^t_{t-(k-1)},Z_{t})\}$ is a Markov process,
where $\{S_{t}\}$ has a state space
$\mathcal{S}=\mathcal{A}^k\times\mathcal{A}$. This is true since
\begin{align*}
Q(S_{t+1}|S^{t})=&Q(X_{t+1},Z_{t+1}|X^{t},Z^{t})\\
                =&Q(X_{t+1}|X^{t},Z^{t})Q(Z_{t+1}|X^{t+1},Z^{t})\\
                =&Q(X_{t+1}|X^{t}_{t-(k-1)})\Pi(X_{t+1},Z_{t+1})\\
                =&Q(S_{t+1}|S_{t}).
\end{align*}
Let $\{x_1^k(i):i=1,\cdots,M^k\}$ denote the set of all possible
$k$-tuples of $\{X_t\}$, and let
$s=(x_1^k(i),z),\bar{s}=(x_1^k(j),\bar{z})$. Then, the transition
matrix $T$ of $\{S_t\}$ has elements $t_{s\bar{s}}\triangleq
Q(S_{t+1}=\bar{s}|S_t=s)=a_{ij}\Pi(x_k(j),\bar{z})$. Since all $A$
that are in $\Theta_{k}^{\delta}$ are irreducible and aperiodic and
$\Pi(x_k(j),\bar{z})>0$, $\forall x_k(j),\bar{z}$, $T$ is also
irreducible and aperiodic. Hence, $T$ has the unique stationary
distribution $\tau$. Although there are zeros in $T$, by the
construction, any $n$-tuple $s^n$ has positive probability. Since
$\{S_t\}$ is also stationary, we have
$$
Q(S^n=s^n)=\tau_{s_1}\prod_{t=k}^{n-1}t_{s_{t}s_{t+1}}=\tau_{s_1}\prod_{(s,\bar{s})}t_{s\bar{s}}^{n_{s\bar{s}}},
$$
where
$$
n_{s\bar{s}}\triangleq\sum_{t=k}^{n-1}\mathbf{1}(S_t=s,S_{t+1}=\bar{s}).
$$
For another probability law $Q^{'}\in\Theta_k^{\delta}$, we have
\begin{align}
 &|\frac{1}{n}\log Q(S^n)-\frac{1}{n}\log Q^{'}(S^n)|\nonumber\\
\leq&|\frac{1}{n}\log\tau_{s_1}-\frac{1}{n}\log\tau^{'}_{s_1}|+|\frac{1}{n}\sum_{(s,\bar{s})}n_{s\bar{s}}\log
t_{s\bar{s}}-\frac{1}{n}\sum_{(s,\bar{s})}n_{s\bar{s}}\log
t^{'}_{s\bar{s}}|\nonumber\\
\leq&|\log\tau_{s_1}-\log\tau^{'}_{s_1}|+\sum_{(s,\bar{s})}|\log
t_{s\bar{s}}-\log t^{'}_{s\bar{s}}|\label{lem2 ineq1}\\
=&|\log\tau_{s_1}-\log\tau^{'}_{s_1}|+\sum_{(i,j)}|\log a_{ij}-\log
a^{'}_{ij}|\label{lem2 ineq2}
\end{align}
where \eq{lem2 ineq1} is from the fact that
$\frac{1}{n}\leq1$,$\frac{n_{s\bar{s}}}{n}\leq1$, and \eq{lem2
ineq2} is from the fact that DMC, $\mathbf{\Pi}$, is equal for $Q$
and $Q^{'}$. The summations are over the pairs that have nonzero
transition probabilities.

Since the function $f(x)=\log x$ is a uniformly continuous function
for $\delta\leq x<1$, and $a_{ij}\geq\delta$ that occur in the
summation, we have for $\epsilon>0$,
$$
\sum_{(i,j)}|\log a_{ij}-\log a^{'}_{ij}|<\frac{\epsilon}{2} \quad
\textrm{if}\quad \|Q-Q^{'}\|_{1}<\delta_1(\epsilon).
$$
Also, we know that all the elements of the stationary distribution
of $T$ are bounded away from zero, since the largest element of the
stationary distribution of $T$ is lower bounded by
$\frac{1}{M^{k+1}}$, and any state can be reached by finite number
of steps whose transition probabilities are bounded away from zero.
Therefore, for some $C_1<\infty$, we have,
$$
|\log\tau_{s_1}-\log\tau^{'}_{s_1}|<C_1|\tau_{s_1}-\tau^{'}_{s_1}|.
$$
Then, from the result of the sensitivity of the stationary
distribution of a Markov chain \cite{GolubMeyer86}, for some
$C_2<\infty$, we have,
$$
|\tau_{s1}-\tau^{'}_{s1}|\leq
C_2\sum_{(s,\bar{s})}|t_{s\bar{s}}-t^{'}_{s\bar{s}}|=C_2\sum_{(i,j)}|a_{ij}-a^{'}_{ij}|.
$$
Hence, for $\epsilon>0$, we obtain,
$$
|\log\tau_{s_1}-\log\tau^{'}_{s_1}|<\frac{\epsilon}{2}
\quad\textrm{if} \quad \|Q-Q^{'}\|_{1}<\delta_2(\epsilon).
$$
Therefore, by letting
$\delta(\epsilon)=\min(\delta_1(\epsilon),\delta_2(\epsilon))$, we
have
$$
\left|\frac{1}{n}\log Q(S^n)-\frac{1}{n}\log
Q^{'}(S^n)\right|<\epsilon \quad \textrm{if}\quad
\|Q-Q^{'}\|_{1}<\delta(\epsilon).
$$

Let us now go back to the original process $Z$. From
$$
\left|\frac{1}{n}\log Q(S^n)-\frac{1}{n}\log
Q^{'}(S^n)\right|<\epsilon,
$$
we have
$$
Q^{'}(X^n,Z^n)< \exp(n\epsilon)Q(X^n,Z^n),
$$
thus,
\begin{align*}
Q^{'}(Z^n)=&\sum_{x^n}Q^{'}(x^n,Z^n)<
          \exp(n\epsilon)\sum_{x^n}Q(x^n,Z^n)\\
          =&\exp(n\epsilon)Q(Z^n)
\end{align*}
where the summations are again over the sequences that have nonzero
probabilities. By changing the role of $Q$, and $Q^{'}$, we get the
result that $\frac{1}{n}\log Q(Z^n)$ is an equicontinuous sequence.
Therefore, we have the uniform convergence of the lemma.\quad
$\blacksquare$

\begin{lem}\emph{(}Uniform Concentration\emph{)}\label{lemma for concentration}
Suppose $Q\in\Theta_k^{\delta}$ for some fixed $\delta>0$. Let
$\hat{\mathbf{X}}_{Q}^{\epsilon}$ be the randomized filter defined
in \eq{randomized filter}. Then,
$$
\lim_{n\rightarrow\infty}\Big(L_{\hat{\mathbf{X}}_Q^{\epsilon}}(X^{n},Z^{n})-E\Big(L_{\hat{\mathbf{X}}_Q^{\epsilon}}(X^{n},Z^{n})\Big)\Big)=0
\quad\textrm{\emph{a.s.}\quad uniformly on $\Theta_{k}^{\delta}$}
$$
\end{lem}
\emph{Proof:} This lemma shows the uniform concentration property of
$L_{\hat{\mathbf{X}}_Q^{\epsilon}}(X^{n},Z^{n})$. The randomization
of the filter is needed to deal with ties occur in deciding the
Bayes response. A detailed proof of this lemma is given in Appendix
2.

\begin{lem}\emph{(}Continuity\emph{)}\label{lemma for difference of expected loss}
Consider a single letter filtering setting. Suppoes $Q$ is some
other joint probability law of $X$ and $Z$. Define single letter
filters $\hat{X}_{P}(z)$ and $\hat{X}_{Q}^{\epsilon}(z)$ as
\begin{align*}
\hat{X}_{P}(z)[\hat{x}]=&Pr(B(\mathbf{P}_{X|z})=\hat{x})\\
\hat{X}_{Q}^{\epsilon}(z)[\hat{x}]=&Pr(B(\mathbf{Q}_{X|z}+\mathbf{U})=\hat{x}),
\end{align*}
where $\mathbf{U}\in\mathbb{R}^M$ is a uniform random vector in
$B_{\epsilon}$ as before. Then,
$$
E\Big(\ell(X,\hat{X}_{Q}^{\epsilon}(Z))\Big)-E\Big(\ell(X,\hat{X}_{P}(Z))\Big)
\leq\Lambda_{\max}K_{\mathbf{\Pi}}\cdot
\|\mathbf{P}_Z-\mathbf{Q}_Z\|_{1}+C_{\mathbf{\Lambda}}\cdot\epsilon,
$$
where the expectations on the left hand side of the inequality are
under $P$ and $K_{\mathbf{\Pi}}=\sum_{i=1}^M\|\Pi_i^{-1}\|_{2}$, and
$C_{\mathbf{\Lambda}}=\max_{a,b\in\mathcal{A}}\|\mathbf{\lambda}_a-\mathbf{\lambda}_b\|_2$.
\end{lem}

 This lemma states that the excess expected loss of a
randomized filter optimized for a mismatched probability law can be
upper bounded by the $L_1$ difference between the true and the
mismatched probability laws of output symbol, plus a small constant
term which diminishes with the randomization probability. This is
somewhat analogous to a  for the prediction which was derived in
\cite[(20)]{Merhav1998}.

\emph{Proof of Lemma \ref{lemma for difference of expected loss}:}
Define $\hat{X}_{Q}(z)[\hat{x}]=Pr(B(\mathbf{Q}_{X|z})=\hat{x}).$
Then,
\begin{align}
 &E\Big(\ell(X,\hat{X}_{Q}^{\epsilon}(Z))\Big)-E\Big(\ell(X,\hat{X}_{P}(Z))\Big)\nonumber\\
=&\sum_{x,z}P(x,z)\Big(\ell(x,\hat{X}_{Q}^{\epsilon}(z))-\ell(x,\hat{X}_{P}(z))\Big)\nonumber\\
\leq&\sum_{x,z}\Big(Q(x,z)+|P(x,z)-Q(x,z)|\Big)\Big(\ell(x,\hat{X}_{Q}(z))-\ell(x,\hat{X}_{P}(z))+\ell(x,\hat{X}_{Q}^{\epsilon}(z))-\ell(x,\hat{X}_{Q}(z))\Big)\nonumber\\
\leq&\sum_{x,z}|P(x,z)-Q(x,z)|\cdot\Big(\ell(x,\hat{X}_{Q}(z))-\ell(x,\hat{X}_{P}(z))\Big)\label{lem4 ineq7}\\
 &+\sum_{x,z}\Big(Q(x,z)+|P(x,z)-Q(x,z)|\Big)\cdot\Big(\ell(x,\hat{X}_{Q}^{\epsilon}(z))-\ell(x,\hat{X}_{Q}(z))\Big)\nonumber\\
=&\sum_{x,z}|P(x,z)-Q(x,z)|\cdot\Big(\ell(x,\hat{X}_{Q}^{\epsilon}(z))-\ell(x,\hat{X}_{P}(z))\Big)+\sum_{x,z}Q(x,z)\Big(\ell(x,\hat{X}_{Q}^{\epsilon}(z))-\ell(x,\hat{X}_{Q}(z))\Big)\label{lem4 ineq8}\\
\leq&\Lambda_{\textit{max}}\sum_{x,z}|P(x,z)-Q(x,z)|+\sum_{x,z}Q(x,z)\Big(\ell(x,\hat{X}_{Q}^{\epsilon}(z))-\ell(x,\hat{X}_{Q}(z))\Big),\label{lem4
ineq6}
\end{align}
where \eq{lem4 ineq7} is from the fact that
$\sum_{x,z}Q(x,z)(\ell(x,\hat{X}_{Q}(z))-\ell(x,\hat{X}_{P}(z)))\leq0$
and \eq{lem4 ineq8} is from rearranging terms in the summation. Now,
let's bound the first term in \eq{lem4 ineq6}.
\begin{align}
 &\Lambda_{\max}\sum_{x,z}|P(x,z)-Q(x,z)|\nonumber\\
=&\Lambda_{\max}\sum_{x}|P(x)-Q(x)|\Big(\sum_{z}\Pi(x,z)\Big)\nonumber\\
=&\Lambda_{\max}\sum_{x}|P(x)-Q(x)|\label{lem4 ineq1}\\
=&\Lambda_{\max}\sum_{i}|(\mathbf{P}_Z-\mathbf{Q}_Z)^T\Pi_{i}^{-1}|\nonumber\\
\leq&\Lambda_{\max}\sum_{i}\|\Pi_{i}^{-1}\|_{2}\cdot\|\mathbf{P}_Z-\mathbf{Q}_Z\|_{2}\label{lem4 ineq3}\\
\leq&\Lambda_{\max}K_{\mathbf{\Pi}}\cdot\|\mathbf{P}_Z-\mathbf{Q}_Z\|_{1},\label{lem4
ineq13}
\end{align}
where \eq{lem4 ineq1} is from the fact that $\sum_z\Pi(x,z)=1$,
\eq{lem4 ineq3} is from Cauchy-Schwartz inequality, and \eq{lem4
ineq13} is from the fact that $L_2$-norm is less than or equal to
$L_1$-norm.

The second term in \eq{lem4 ineq6} becomes
\begin{align}
 &\sum_{x,z}Q(x,z)\Big(\ell(x,\hat{X}_{Q}^{\epsilon}(z))-\ell(x,\hat{X}_{Q}(z))\Big)\nonumber\\
=&\sum_{z}Q(z)\sum_{x}Q(x|z)\sum_{\hat{x}}\Lambda(x,\hat{x})\cdot\Big(\hat{X}_{Q}^{\epsilon}(z)[\hat{x}]-\hat{X}_{Q}(z)[\hat{x}]\Big)\nonumber\\
=&\sum_{z}Q(z)\sum_{\hat{x}}\Big(\hat{X}_{Q}^{\epsilon}(z)[\hat{x}]-\hat{X}_{Q}(z)[\hat{x}]\Big)\sum_{x}\Lambda(x,\hat{x})Q(x|z)\nonumber\\
=&\sum_{z}Q(z)\sum_{\hat{x}}\Big(\hat{X}_{Q}^{\epsilon}(z)[\hat{x}]-\hat{X}_{Q}(z)[\hat{x}]\Big)\cdot\lambda_{\hat{x}}^T\mathbf{Q}_{X|z}.\label{lem4
ineq4}
\end{align}
It is easy to see that the inner summation in \eq{lem4 ineq4} is
always nonnegative since by definition, $\hat{X}_Q(z)$ assigns
probability 1 to $B(\mathbf{Q}_{X|z})$. Now, for a given $Q$, define
\begin{align}
\mathbf{U}_{\max}=&\arg\max_{\mathbf{U}\in
B_{\epsilon}}\Big(\lambda_{B(\mathbf{Q}_{X|z}+\mathbf{U})}-\lambda_{B(\mathbf{Q}_{X|z})}\Big)^T\mathbf{Q}_{X|z},\label{lem4
ineq14}
\end{align}
resolving ties arbitrarily. Then, we have,
\begin{align}
 &\sum_{\hat{x}}\Big(\hat{X}_{Q}^{\epsilon}(z)[\hat{x}]-\hat{X}_{Q}(z)[\hat{x}]\Big)\cdot\mathbf{\lambda}_{\hat{x}}^T\mathbf{Q}_{X|z}\nonumber\\
=&\Big(\sum_{\hat{x}}\Big(\hat{X}_{Q}^{\epsilon}(z)[\hat{x}]\cdot\mathbf{\lambda}_{\hat{x}}\Big)-\mathbf{\lambda}_{B(\mathbf{Q}(X|z))}\Big)^T\mathbf{Q}_{X|z}\nonumber\\
\leq&\Big(\mathbf{\lambda}_{B(\mathbf{Q}(X|z)+\mathbf{U}_{\max})}-\mathbf{\lambda}_{B(\mathbf{Q}(X|z))}\Big)^T\mathbf{Q}_{X|z}\label{lem4 ineq11}\\
\leq&\Big(\mathbf{\lambda}_{B(\mathbf{Q}(X|z))}-\mathbf{\lambda}_{B(\mathbf{Q}_{X|z}+\mathbf{U}_{\max})}\Big)^T\mathbf{U}_{\max}\label{lem4 ineq12}\\
\leq&\max_{a,b\in\mathcal{A}}\|\mathbf{\lambda}_a-\mathbf{\lambda}_b\|_{2}\cdot\|\mathbf{U}_{\max}\|_{2}\label{lem4 ineq5}\\
\leq&C_{\mathbf{\Lambda}}\cdot\epsilon,\nonumber
\end{align}
where \eq{lem4 ineq11} follows from \eq{lem4 ineq14}, \eq{lem4
ineq12} follows from the fact
$$
\lambda_{B(\mathbf{Q}_{X|z}+\mathbf{U}_{\max})}^T(\mathbf{Q}_{X|z}+\mathbf{U}_{\max})\leq\lambda_{B(\mathbf{Q}_{X|z})}^T(\mathbf{Q}_{X|z}+\mathbf{U}_{\max}),
$$
and \eq{lem4 ineq5} follows from the Cauchy-Schwartz inequality.
Note that depending on $Q$ and $z$, \eq{lem4 ineq11} and \eq{lem4
ineq12} can be both zero and hold with equality. Together with
\eq{lem4 ineq13}, the lemma is proved.\quad $\blacksquare$

Before moving on to Lemma \ref{lemma for two inequalities}, we need
following three definitions. In Lemma \ref{lemma for KL divergence},
we have seen that for $Q\in\Theta_k^{\delta}$, $\mathbf{D}(P_Z\|Q)$
is well-defined. Now, let's consider the case where
$Q\in\Theta_k^{\delta}$ is some function of the noisy observation
$Z^n$ (denoted as $Q[Z^n]$). As mentioned in the footnote of Section
\ref{sec: intuition}, the notion of the relative entropy rate
between $P_Z$ and that random $Q[Z^n]$ is defined in Definition
\ref{random dpq} using Definition \ref{E hat}. Definition \ref{def
for kth order Markov} is also needed for the inequality in Lemma
\ref{lemma for two inequalities}.
\begin{defn}\label{E hat}
Suppose $Q[Z^n]\in\Theta_k^{\delta}$. If $f$ is some function of
$(X^{\infty},Z^{\infty},Q[Z^n])$ such that the expectation
$$
E\Big(f(X^{\infty},Z^{\infty},Q[Z^n])\Big)=\int
f(x^{\infty},z^{\infty},Q[z^n])dP(x^{\infty},z^{\infty})
$$
exists. Then, define the notation $\hat{E}(\cdot)$ as following:
$$
\hat{E}\Big(f(X^{\infty},Z^{\infty},Q[Z^n])\Big)\triangleq\int
f(x^{\infty},z^{\infty},Q[Z^n])dP(x^{\infty},z^{\infty})
$$
That is, in $\hat{E}\Big(f(X^{\infty},Z^{\infty},Q[Z^n])\Big)$, the
Lebesgue integration with respect to the randomness of $Q[Z^n]$ is
excluded.
\end{defn}
\begin{defn}\label{random dpq}
Suppose $Q[Z^n]\in\Theta_k^{\delta}$. Then, the relative entropy
rate between $P_Z$ and $Q[Z^n]$ is defined as,\footnote{Note that
$\mathbf{D}(P_Z\|Q[Z^n])$ is a function of $Z^n$, and still is a
random variable.}
$$
\mathbf{D}(P_Z\|Q[Z^n])\triangleq
\hat{E}\Big(\log\frac{P_Z(Z_0|Z_{-\infty}^{-1})}{Q[Z^n](Z_0|Z_{-\infty}^{-1})}\Big).
$$
\end{defn}
\begin{defn}\label{def for kth order Markov}
Define the $k$-th order Markov approximation of $P_X$ for $n\geq k$
as
$$
P_{X}^{(k)}(X^n)\triangleq
P_X(X^k)\prod_{i=k+1}^{n}P_X(X_{i}|X_{i-k}^{i-1}).
$$
Furthermore, denote $P_Z$ and $P_Z^{(k)}$ as the probability law of
the output of DMC, $\mathbf{\Pi}$, when the probability law of input
is $P_X$ and $P_X^{(k)}$, respectively.\footnote{Here, $P_Z^{(k)}$
is not the $k$-th order Markov approximation of $P_Z$, but is the
distribution of the channel output whose input is $P_X^{(k)}$, the
$k$-th order Markov approximation of the original input distribution
$P_X$.}
\end{defn}
Now, we give following lemma that upper bounds the relative entropy
rate between $P_Z$ and the ML estimator.
\begin{lem}\label{lemma for two inequalities}
For the given sequence $\{\delta_k\}$ defined in Section \ref{sec:
description of the filter} and for fixed $k$, we have
$$
\lim_{n\rightarrow\infty}\mathbf{D}(P_Z\|\hat{Q}_{k,\delta_k}[Z^n])\leq\mathbf{D}(P_X\|P_X^{(k)})\quad\textrm{\emph{a.s.}}
$$
\end{lem}
\emph{Proof:} Recall that $\hat{Q}_{k,\delta_k}[Z^n]$ is an ML
estimator in $\Theta_k^{\delta_k}$ based on the observation $Z^n$.
From \eq{second consistency}, we know that
$$\lim_{n\rightarrow\infty}\mathbf{D}(P_Z\|\hat{Q}_{k,\delta_k}[Z^n])=\min_{Q\in\Theta_k^{\delta_k}}\mathbf{D}(P_Z\|Q)\quad\textrm{\emph{a.s.}}$$
Also, \eq{assumption of source} and Definition \ref{def for kth
order Markov} assures that $P_Z^{(k)}\in\Theta_{k}^{\delta_k}$.
Therefore, we have
$$
\lim_{n\rightarrow
\infty}\mathbf{D}(P_Z\|\hat{Q}_{k,\delta_k}[Z^n])\leq
\mathbf{D}(P_Z\|P_Z^{(k)}) \quad\textrm{\emph{a.s.}}
$$
This is the link where we needed Assumption 1. Now, let's denote
$P^{(k)}$ as the joint probability law of $(X^n,Z^n)$ when the
probability law of input process is $P_X^{(k)}$. Then, by the chain
rule of relative entropy \cite[(2.67)]{CoverThomas91}, we have
\begin{align*}
 &E\Big(\log\frac{P(X^n,Z^n)}{P^{(k)}(X^n,Z^n)}\Big)\\
=&D_n(P_X\|P_X^{(k)})+E\Big(\log\frac{P(Z^n|X^n)}{P^{(k)}(Z^n|X^n)}\Big)\\
=&D_n(P_Z\|P_Z^{(k)})+E\Big(\log\frac{P(X^n|Z^n)}{P^{(k)}(X^n|Z^n)}\Big)
\end{align*}
Since the DMC is fixed, we have
$E\Big(\log\frac{P(Z^n|X^n)}{P^{(k)}(Z^n|X^n)}\Big)=0$. Moreover, by
the nonnegativity of relative entropy,
$E\Big(\log\frac{P(X^n|Z^n)}{P^{(k)}(X^n|Z^n)}\Big)\geq0$.
Therefore, we get $D_n(P_Z\|P_Z^{(k)})\leq D_n(P_X\|P_X^{(k)})$.
Since
$\mathbf{D}(P_X\|P_X^{(k)})=\lim_{n\rightarrow\infty}\frac{1}{n}D_n(P_X\|P_X^{(k)})$
always exists by ergodicity, we have
$$
\mathbf{D}(P_Z\|P_Z^{(k)})\leq\mathbf{D}(P_X\|P_X^{(k)}),
$$
and the lemma is proved.\quad $\blacksquare$



\begin{pfth}\label{proof of the theorem}
\emph{ We are now finally in a position to prove our main theorem.
As mentioned in Section \ref{sec: intuition}, we first fix $k$ and
$\epsilon$, and try to get the inequality in the form of \eq{proof
sketch} to prove Part (a). To refresh, \eq{proof sketch} is given
again here.
\begin{align*}
\limsup_{n\rightarrow\infty}\Big(L_{\hat{\mathbf{X}}_{univ,k}^{\epsilon}}(X^n,Z^n)-\phi_n(P_X,\mathbf{\Pi})\Big)\leq
F\Big(\limsup_{t\rightarrow\infty}\mathbf{D}(P_Z\|Q_k^t),\epsilon\Big)\qquad\textrm{\emph{a.s.}}
\end{align*}
From the definition of
$L_{\hat{\mathbf{X}}_{univ,k}^{\epsilon}}(X^n,Z^n)$,
$$
L_{\hat{\mathbf{X}}_{univ,k}^{\epsilon}}(X^n,Z^n)=\frac{1}{n}\sum_{t=1}^{n}\ell(X_t,\hat{X}_{Q_k^t}^{\epsilon}(Z^t)),
$$
where from \eq{random Q}, we know that $Q_k^t$ is a function of
$Z^{m_i(t)}$. Since $\ell(X_t,\hat{X}_{Q_k^t}^{\epsilon}(Z^t))$ is a
function of $(X_t,Z^t,Q[Z^{m_i(t)}])$, we can define a quantity
$\hat{E}(\ell(X_t,\hat{X}_{Q_k^t}^{\epsilon}(Z^t)))$ from Definition
\ref{E hat}. From this, we also define
$$
\hat{E}\Big(L_{\hat{\mathbf{X}}_{univ,k}^{\epsilon}}(X^n,Z^n)\Big)=\frac{1}{n}\sum_{t=1}^{n}\hat{E}\Big(\ell(X_t,\hat{X}_{Q_k^t}^{\epsilon}(Z^t))\Big).
$$
Now, we have following Corollary \ref{univ concentration} from Lemma
\ref{lemma for concentration}, whose proof is given in Appendix 3.
This corollary is a key step in proving the main theorem, since it
provides a crucial link that enables to get the inequality in
\eq{proof sketch}.
\begin{cor}\label{univ concentration}
For fixed $k$ and $\epsilon$, we have
$$
\lim_{n\rightarrow\infty}\Big(L_{\hat{\mathbf{X}}_{univ,k}^{\epsilon}}(X^n,Z^n)-\hat{E}\Big(L_{\hat{\mathbf{X}}_{univ,k}^{\epsilon}}(X^n,Z^n)\Big)\Big)=0\quad\textrm{a.s.}
$$
\end{cor}
From Corollary \ref{univ concentration}, we have following equality
\begin{align*}
 &\limsup_{n\rightarrow\infty}\Big(L_{\hat{\mathbf{X}}_{univ,k}^{\epsilon}}(X^{n},Z^{n})-
\phi_n(P_X,\Pi)\Big)\\
=&\limsup_{n\rightarrow\infty}\Big(\hat{E}\Big (
L_{\hat{\mathbf{X}}_{univ,k}^{\epsilon}}(X^{n},Z^{n})\Big) -
\phi_n(P_X,\Pi)\Big)\quad\textrm{\emph{a.s.}}
\end{align*}
Therefore , to get the inequality of the form of \eq{proof sketch},
we can equivalently show
$$
\limsup_{n\rightarrow\infty}\Big(\hat{E}\Big (
L_{\hat{\mathbf{X}}_{univ,k}^{\epsilon}}(X^{n},Z^{n})\Big) -
\phi_n(P_X,\Pi)\Big)\leq
F\Big(\limsup_{t\rightarrow\infty}\mathbf{D}(P_Z\|Q_k^t),\epsilon\Big).
$$
Now, let's consider following chain of inequalities:
\begin{align}
 &\hat{E}\Big ( L_{\hat{\mathbf{X}}_{univ,k}^{\epsilon}}(X^{n},Z^{n})\Big) - \phi_n(P_X,\Pi)\nonumber \\
=&\frac{1}{n}\sum_{t=1}^{n}\Big(\hat{E}\Big(\ell(X_{t},\hat{X}_{Q_k^t}^{\epsilon}(Z^{t}))\Big)-\hat{E}\Big(\ell(X_{t},\hat{X}_{P}(Z^{t}))\Big)\Big)\quad\textrm{\emph{a.s.}}\nonumber\\
=&\frac{1}{n}\sum_{t=1}^{n}\hat{E}\Big(\hat{E}\Big(\ell(X_{t},\hat{X}_{Q_k^t}^{\epsilon}(Z_{t},Z^{t-1}))|Z^{t-1}\Big)-\hat{E}\Big(\ell(X_{t},\hat{X}_{P}(Z_{t},Z^{t-1}))|Z^{t-1}\Big)\Big)\quad\textrm{\emph{a.s.}}\nonumber\\
\leq& \frac{K_{\mathbf{\Pi}}\Lambda_{\max}}{n}
\sum_{t=1}^{n}\hat{E}\|\mathbf{P}_{Z_t|Z^{t-1}}-\mathbf{Q_k^{t}}_{Z_t|Z^{t-1}}\|_{1}+C_{\mathbf{\Lambda}}\cdot\epsilon\quad\textrm{\emph{a.s.}}\label{pfth ineq1}\\
\leq&\frac{\sqrt{2\ln2}K_{\mathbf{\Pi}}\Lambda_{\max}}{n}
\sum_{t=1}^{n}\hat{E}\sqrt{\hat{E}\Big(\log\frac{P(Z_t|Z^{t-1})}{Q_k^t(Z_t|Z^{t-1})}\Big|Z^{t-1}\Big)}+C_{\mathbf{\Lambda}}\cdot\epsilon\quad\textrm{\emph{a.s.}}\label{pfth ineq2}\\
\leq&\sqrt{2\ln2}K_{\mathbf{\Pi}}\Lambda_{\max}\sqrt{\frac{1}{n}\sum_{t=1}^{n}\hat{E}\Big(\log\frac{P(Z_t|Z^{t-1})}{Q_{k}^{t}(Z_t|Z^{t-1})}\Big)}+C_{\mathbf{\Lambda}}\cdot\epsilon.\quad\textrm{\emph{a.s.}}\label{pfth
ineq3}
\end{align}
\eq{pfth ineq1} is obtained from Lemma \ref{lemma for difference of
expected loss}, since $\mathbf{\Pi}$ does not vary with $t$, and
given $Z^{t-1}$, estimating $X_t$ based on $Z^t$ is equivalent to
the single letter setting as in Lemma \ref{lemma for difference of
expected loss} with the corresponding conditional distribution.
Also, \eq{pfth ineq2} is from Pinsker's inequality, and \eq{pfth
ineq3} is from Jensen's inequality. By taking $\limsup$ on both
sides, we have
\begin{align*}
 &\limsup_{n\rightarrow\infty}\Big(\hat{E}\Big(L_{\hat{\mathbf{X}}_{univ,k}^{\epsilon}}(X^{n},Z^{n})\Big)-
\phi_n(P_X,\Pi)\Big)\\
\leq&\sqrt{2\ln2}K_{\mathbf{\Pi}}\Lambda_{\max}\sqrt{\limsup_{n\rightarrow\infty}\frac{1}{n}\sum_{t=1}^{n}\hat{E}\Big(\log\frac{P(Z_t|Z^{t-1})}{Q_{k}^{t}(Z_t|Z^{t-1})}\Big)}+C_{\mathbf{\Lambda}}\cdot\epsilon\quad\textrm{\emph{a.s.}}
\end{align*}
since the square root function is a continuous function. For the
expression inside the square root of the right-hand side of the
inequality,
\begin{align}
&\limsup_{n\rightarrow\infty}\frac{1}{n}\sum_{t=1}^{n}\hat{E}\Big(\log\frac{P(Z_t|Z^{t-1})}{Q_{k}^{t}(Z_t|Z^{t-1})}\Big)\nonumber\\
=&\limsup_{t\rightarrow\infty}\hat{E}\Big(\log\frac{P(Z_t|Z^{t-1})}{Q_{k}^{t}(Z_t|Z^{t-1})}\Big)\quad\textrm{\emph{a.s.}}\label{pfth ineq4}\\
=&\limsup_{t\rightarrow\infty}\hat{E}\Big(\log\frac{P(Z_0|Z_{-\infty}^{-1})}{Q_{k}^{t}(Z_0|Z_{-\infty}^{-1})}\Big)\quad\textrm{\emph{a.s.}}\label{pfth ineq5}\\
=&\limsup_{t\rightarrow\infty}\mathbf{D}(P_Z\|Q_k^t)\quad\textrm{\emph{a.s.}}\label{pfth
ineq6}
\end{align}
where \eq{pfth ineq4} is from Ces\'aro's mean convergence theorem;
\eq{pfth ineq5} is from the fact that $P(Z_0|Z_{-t}^{-1})\rightarrow
P(Z_0|Z_{-\infty}^{-1})$ almost surely by martingale convergence
theorem, and $Q_{k}^{t}(Z_t|Z^{t-1})\rightarrow
Q_{k}^{t}(Z_0|Z_{-\infty}^{-1})$ almost surely by Lemma \ref{lemma
for uniform convergence}, and \eq{pfth ineq6} is from Definition
\ref{random dpq}. Therefore,
\begin{align}
 &\limsup_{n\rightarrow\infty}\Big(\hat{E}\Big(L_{\hat{\mathbf{X}}_{univ,k}^{\epsilon}}(X^{n},Z^{n})\Big)-\phi_n(P_X,\Pi)\Big)\nonumber\\
 &\limsup_{n\rightarrow\infty}\Big(L_{\hat{\mathbf{X}}_{univ,k}^{\epsilon}}(X^{n},Z^{n})-\phi_n(P_X,\Pi)\Big)\nonumber\\
\leq&2\sqrt{2\ln2}K_{\Pi}\Lambda_{max}\sqrt{\limsup_{t\rightarrow\infty}\mathbf{D}(P_Z\|Q_k^t)}+C_{\mathbf{\Lambda}}\cdot\epsilon\quad\textrm{\emph{a.s.}}\label{actual
F}
\end{align}
which finally is in the form of \eq{proof sketch}. Now, we need to
check if the right-hand side of \eq{actual F} goes to zero if we let
$k\rightarrow\infty$ and $\epsilon\downarrow0$. To see this,
consider following further upper bounds.
\begin{align}
 &\limsup_{t\rightarrow\infty}\mathbf{D}(P_Z\|Q_k^t)\nonumber\\
=&\limsup_{t\rightarrow\infty}\mathbf{D}(P_Z\|\hat{Q}_{k,\delta_k}[Z^{t}])\label{pfth
ineq7}\\
\leq&\mathbf{D}(P_X\|P_{X_k})\label{pfth ineq8},
\end{align}
where \eq{pfth ineq7} is from the fact that
$m_{i(t)}\rightarrow\infty$ as $t\rightarrow\infty$, and \eq{pfth
ineq8} is from Lemma \ref{lemma for two inequalities}. The
inequality \eq{pfth ineq8} holds for every $k$, and by
Shannon-McMillan-Breiman Theorem, we know
$\mathbf{D}(P_X\|P_{X_{k}})\rightarrow 0$ as $k\rightarrow \infty$.
Therefore,
$$
\lim_{k\rightarrow\infty}\limsup_{t\rightarrow\infty}\mathbf{D}(P_Z\|Q_k^t)=0,
$$
and thus,
$$
\lim_{k\rightarrow\infty}\limsup_{n\rightarrow\infty}\Big(L_{\hat{\mathbf{X}}_{univ,k}^{\epsilon}}(X^{n},Z^{n})-\phi_n(P_X,\Pi)\Big)\leq
C_{\mathbf{\Lambda}}\cdot\epsilon \quad\textrm{\emph{a.s.}}
$$
Finally, sending $\epsilon$ to zero, Part (a) of the theorem is
proved. Part (b) follows directly from (a), and Reverse Fatou's
Lemma. That is,
\begin{align*}
&\lim_{k\rightarrow\infty}\limsup_{n\rightarrow\infty}\Big(E\Big(L_{\hat{\mathbf{X}}_{univ,k}^{\epsilon}}(X^{n},Z^{n})\Big)-\phi_n(P_X,\Pi)\Big)\\
=&\lim_{k\rightarrow\infty}\limsup_{n\rightarrow\infty}E\Big(L_{\hat{\mathbf{X}}_{univ,k}^{\epsilon}}(X^{n},Z^{n})-\phi_n(P_X,\Pi)\Big)\\
\leq&\lim_{k\rightarrow\infty}E\Bigg(\limsup_{n\rightarrow\infty}\Big(L_{\hat{\mathbf{X}}_{univ,k}^{\epsilon}}(X^{n},Z^{n})-\phi_n(P_X,\Pi)\Big)\Bigg)\\
\leq&C_{\mathbf{\Lambda}}\cdot\epsilon
\end{align*}
Note that the expectation here is with respect to the randomness of
probability law within the paranthesis, too. By sending $\epsilon$
to zero, Part (b) is proved.\quad$\blacksquare$}
\end{pfth}

\section{Extension: Universal filtering for channel with memory}\label{sec: extension}
Now, let's extend our result to the case where channel has memory.
With the identical assumption on $\{X_t\}$, now suppose $\{Z_t\}$ is
expressed as
\begin{align}
Z_t=X_t\oplus N_t\label{ext eq1}
\end{align}
where $\oplus$ denotes modulo-$M$ addition, and $\{N_t\}$ is an
$\mathcal{A}$-valued noise process which is not necessarily
memoryless. We assume we have a complete knowledge of the
probability law of $\{N_t\}$. Specifically, let's consider the case
where $\{N_t\}$ is FS-HMP, that is, it is an output of an invertible
memoryless channel
$\mathbf{\Gamma}=\{\Gamma(i,j)\}_{i,j\in\mathcal{A}}$ whose input is
irreducible, aperiodic $\ell$-th order Markov chain, $\{S_t\}$,
which is independent of $\{X_t\}$. Let
$\Gamma_{\min}=\min_{i,j\in\mathcal{A}}\{\Gamma(i,j)\}$, and suppose
$\Gamma_{\min}>0$. For simplicity, assume that the alphabet size of
$\{S_t\}$ is also $\mathcal{A}$.

In this model, the channel between $X_t$ and $Z_t$ at time $t$ is an
$M$-ary symmetric channel, which is specified by the $S_t$-th row of
$\mathbf{\Gamma}$. Let's define an $M\times M$ matrix
$\mathbf{\Pi}_t$ whose $(x_t,z_t)$-th element is
\begin{align*}
\Pi_t(x_t,z_t)=&P_{N_t}(z_t\ominus x_t)\\
            =&Pr(Z_t=z_t|X_t=x_t)\\
            =&\sum_{s_t}Pr(Z_t=z_t|X_t=x_t,S_t=s_t)Pr(S_t=s_t),
\end{align*}
where $\ominus$ denotes modulo-$M$ subtraction. Now, let's make
following assumptions on the noise process.
\begin{itemize}
\item $\{N_t\}$ is stationary, i.e., $\mathbf{\Pi}_t$ is identical for
$\forall t$
\item $\mathbf{\Pi}_t$ is invertible
\item $\exists \alpha$\textrm{       such that}\quad $Pr(S_t|S_{t-\ell}^{t-1})\geq \alpha>0, \quad \textrm{for} \quad\forall S^t_{t-\ell}(
\omega)$ \label{ext eq2}
\end{itemize}
As stated in \cite[2-A]{zhang05}, the first and the second
assumptions are rather benign. Especially, for the second
assumption, it can be shown that under benign conditions on the
parametrization, almost all parameter values except for those in a
set of Lebesgue measure zero, give rise to a process satisfying this
assumption. Also, since this only corresponds to the case when $k=0$
in \cite[Assumption 1]{zhang05}, it is a much weaker assumption. The
third assumption is a similar positivity assumption as Assumption 1,
which enables our universal filtering scheme.

Under these assumptions on the noise process, we can extend our
scheme to do the universal filtering for this channel. First, we can
convert this channel to the equivalent memoryless channel,
$\mathbf{\Xi}=\{\xi((i,j),h)\}_{i,j,h\in\mathcal{A}}$ , where the
input process is $\{(X_t,S_t)\}$ and the output is $\{Z_t\}$. Here,
$\mathbf{\Xi}$ is $M^2\times M$ matrix, and the channel transition
probability is
$$
\xi((i,j),h)=\Gamma(j,h\ominus i)\quad\textrm{$\forall i,j,k$}.
$$
To do the filtering, we apply our scheme to this equivalent
memoryless channel. For fixed $k\geq \ell$, as in Section
\ref{Hidden Markov Processes}, define a parameter set of HMPs,
$\Theta_k$, whose Markov chain has $M^{k+\ell}$ states, and the
memoryless channel is $\mathbf{\Xi}$. The $k$-th order conditional
probability of our new input process is
\begin{align}
 &Pr(X_t,S_t|X^{t-1}_{t-k},S^{t-1}_{t-k})\nonumber\\
=&Pr(X_t|X^{t-1}_{t-k})\cdot Pr(S_t|S^{t-1}_{t-\ell})\nonumber\\
\geq&\delta_k\cdot\alpha.\label{ext eq3}
\end{align}
where \eq{ext eq3} is from Assumption 1 and the third condition on
the noise process. Let $\gamma_k=\delta_k\cdot\alpha$. Then, we can
model $\{Z_t\}$ in $\Theta_k^{\gamma_k}$, or equivalently, model
$(X_t,S_t)$ as $k$-th order Markov chain, and obtain $Q_k^t$, the ML
estimator in $\Theta_k^{\gamma_k}$ based on $Z^{m_{i(t)}}$. By
forward recursion, we can get $Q_k^t(X_t,S_t|Z^t)$, and by summing
over $S_t$'s we can calculate $\mathbf{Q}_{kX_t|Z^t}^t$. Then,
finally we define our sequence of universal filtering schemes as,
$$
\hat{\mathbf{X}}_{univ,k}^{\epsilon}=\{\hat{X}_{Q_k^t,t}^{\epsilon}\},
$$
exactly the same as we proposed in Section \ref{sec: description of
the filter}.

The analysis of this scheme is identical to the one given in the
proof of the main theorem. \eq{pfth ineq1}, which is the only place
where the invertibility of the $\mathbf{\Pi}$ is used, can also be
obtained in this case due to the second assumption of the noise
process. Thus, we again get
\begin{align*}
 &\limsup_{n\rightarrow\infty}\Big(L_{\hat{\mathbf{X}}_{univ,k}^{\epsilon}}(X^{n},Z^{n})-\phi_n(P_X,\Pi)\Big)\\
\leq&2\sqrt{2\ln2}K_{\Pi}\Lambda_{max}\sqrt{\limsup_{t\rightarrow\infty}\mathbf{D}(P_Z\|Q_k^t)}+C_{\mathbf{\Lambda}}\cdot\epsilon\quad\textrm{\emph{a.s.}}
\end{align*}
Since
$$
\limsup_{t\rightarrow\infty}\mathbf{D}(P_Z\|Q_k^t)=\limsup_{t\rightarrow\infty}\mathbf{D}(P_Z\|\hat{Q}_{k,\gamma_k}[Z^t])\leq\mathbf{D}(P_X\|P_{X_k})
$$
by the same argument as Lemma \ref{lemma for two inequalities}, we
have the same result as Theorem 1. Thus, we can successfully extend
our scheme to the case where the channel noise is FS-HMP with some
mild assumptions.

\section{Conclusion and future work}\label{sec: conclusion}
In this paper, we proved that, for the known, invertible DMC, a
family of filters based on HMPs is universally asymptotically
optimal for any general stationary and ergodic $\{X_t\}$ satisfying
some mild positivity condition. That is, we showed that our sequence
of schemes indexed by $k$ and $\epsilon$ achieves the best
asymptotically optimal performance regardless of clean source
distribution. We could also extend this scheme to the case where
channel has memory, especially where the channel noise process is
FS-HMP. The future direction of the work would be to ascertain the
relationship between $k$ and $n$, such that we can devise a single
scheme that grows $k$ with some rate related to $n$. Attempting to
loosen the positivity assumption that we made in our main theorem
and extending our discrete universal filtering schemes to discrete
universal denoising schemes are additional future directions of our
research.

\section*{Acknowledgement}
The authors are grateful to Erik Ordentlich and Sergio Verd\'u for
helpful discussions.

\section*{Appendix 1}\label{appendix for uniform convergence}
Here, we revise three lemmas from \cite{BaumPetrie66} regarding
probability law of HMP. These are needed to prove Lemma \ref{lemma
for uniform convergence}. For the following three lemmas, fix $k$
and $\delta$, and suppose $Q\in\Theta_{k}^{\delta}$. Also, fix some
$m\in\mathbb{N}$, such that $m\geq k$. Proofs are similar to
\cite[Appendix]{BaumPetrie66}. Note that $\{X_t\}$ is still our
clean signal and $\{Z_t\}$ is the noisy observed signal (not
necessarily a HMP).

\begin{lem}\label{first lemma for uniform convergence}
\emph{We have$$
Q(X_{t+m}=j|X_{t}=i,Z_{-\infty}^{\infty})\geq\mu_{\delta,k,m},
$$
where
$\mu_{\delta,m,k}=(1+\frac{M-1}{(\delta\cdot\Pi_{\min})^{m+k}})^{-1}$
is independent of $Q,Z_{-\infty}^{\infty},i,j$.}
\end{lem}

\emph{Proof:}
\begin{align}
&\frac{Q(X_{t+m}=j|X_{t}=i,Z_{-\infty}^{\infty})}{Q(X_{t+m}=j^{'}|X_{t}=i,Z_{-\infty}^{\infty})}\nonumber\\
=&\frac{Q(X_{t+m}=j,Z_{-\infty}^{\infty}|X_{t}=i)}{Q(X_{t+m}=j^{'},Z_{-\infty}^{\infty}|X_{t}=i)}\nonumber\\
=&\frac{Q(X_{t+m}=j,Z_{t+m+k+1}^{\infty}|X_{t}=i)}{Q(X_{t+m}=j^{'},Z_{t+m+k+1}^{\infty}|X_{t}=i)}\cdot\frac{Q(Z_{t+1}^{t+m+k}|X_t=i,X_{t+m}=j)}
{Q(Z_{t+1}^{t+m+k}|X_t=i,X_{t+m}=j^{'})}\label{l6 eq1}
\end{align}
Now, let's bound the terms in \eq{l6 eq1}. First,
\begin{align*}
 &\frac{Q(X_{t+m}=j,Z_{t+m+k+1}^{\infty}|X_{t}=i)}{Q(X_{t+m}=j^{'},Z_{t+m+k+1}^{\infty}|X_{t}=i)}\\
=&\frac{\sum_{j_{0}}Q(X_{t+m+k}=j_{0},X_{t+m}=j,Z_{t+m+k+1}^{\infty}|X_{t}=i)}{\sum_{j_{0}}Q(X_{t+m+k}=j_{0},X_{t+m}=j^{'},Z_{t+m+k+1}^{\infty}|X_{t}=i)}\\
=&\frac{\sum_{j_{0}}a_{ij}^{m}a_{jj_{0}}^{k}Q(Z_{t+m+k+1}^{\infty}|X_{t+m+k}=j_{0})}{\sum_{j_{0}}a_{ij^{'}}^{m}a_{j^{'}j_{0}}^{k}Q(Z_{t+m+k+1}^{\infty}|X_{t+m+k}=j_{0})}.
\end{align*}
Note that $a_{ij}^m\geq \delta^m$ and $a_{jj_0}^k\geq\delta^k$,
$\forall i,j,j_0$ from the assumption of $\Theta_k^{\delta}$. Let
$Q(Z_{t+m+k+1}^{\infty}|X_{t+m+k}=j_{0})=\alpha_{j_{0}}$. Then, the
last expression is
\begin{align}
 &\frac{a_{ij}^m}{a_{ij^{'}}^m}\frac{\sum_{j_{0}}a_{jj_{0}}^k\alpha_{j_{0}}}{\sum_{j_{0}}a_{j^{'}j_{0}}^k\alpha_{j_{0}}}.\label{l6
 eq2}
\end{align}
Since
$$
\frac{\sum_{j_{0}}a_{jj_{0}}^k\alpha_{j_{0}}}{\sum_{j_{0}}a_{j^{'}j_{0}}^k\alpha_{j_{0}}}=\frac{\sum_{j_{0}}\alpha_{j_{0}}a_{j^{'}j_{0}}^k\frac{a_{jj_{0}}^k}{a_{j^{'}j_{0}}^k}}{\sum_{j_{0}}\alpha_{j_{0}}a_{j^{'}j_{0}}^k}
\leq\max_{j_{0}}\Big(\frac{a_{jj_{0}}^k}{a_{j^{'}j_{0}}^k}\Big),
$$
we have
\begin{align}
\eq{l6
eq2}\leq\frac{a_{ij}^m}{a_{ij^{'}}^m}\max_{j_{0}}\Big(\frac{a_{jj_{0}}^k}{a_{j^{'}j_{0}}^k}\Big)
\leq
\max_{i,j,j^{'},j_{0}}\Big(\frac{a_{ij}^{m}a_{jj_{0}}^k}{a_{ij^{'}}^{m}a_{j^{'}j_{0}}^k}\Big)
\leq\frac{1}{\delta^{m+k}}.\label{l6 eq4}
\end{align}
Now let's look at the second term in \eq{l6 eq1}. That is,
\begin{align}
 &\frac{Q(Z_{t+1}^{t+m+k}|X_t=i,X_{t+m}=j)} {Q(Z_{t+1}^{t+m+k}|X_t=i,X_{t+m}=j^{'})}\nonumber\\
=&\frac{\sum_{x_{\mathcal{T}}}Q(Z_{t+1}^{t+m+k}|X_t=i,X_{t+m}=j,X_{\mathcal{T}}=x_{\mathcal{T}})\cdot
Q(X_{\mathcal{T}}=x_{\mathcal{T}}|X_t=i,X_{t+m}=j)}
{\sum_{x_{\mathcal{T}}}Q(Z_{t+1}^{t+m+k}|X_t=i,X_{t+m}=j^{'},X_{\mathcal{T}}=x_{\mathcal{T}})\cdot
Q(X_{\mathcal{T}}=x_{\mathcal{T}}|X_t=i,X_{t+m}=j^{'})}\nonumber\\
\leq&\frac{1}{(\Pi_{\min})^{m+k}}\label{l6 eq3}
\end{align}
where $\mathcal{T}=\{t+1,\cdots,t+m+k\} \backslash \{t,t+m\}$. Thus,
from \eq{l6 eq4} and \eq{l6 eq3},
$$
\eq{l6 eq1}\leq\frac{1}{(\delta\cdot\Pi_{\min})^{m+k}}.
$$
Let now $\rho_{j}\triangleq
Q(X_{t+m}=j|X_{t}=i,Z_{-\infty}^{\infty})$, then
$1=\rho_j+\sum_{j^{'}\neq
j}\rho_{j^{'}}\leq\rho_{j}+(M-1)\frac{\rho_{j}}{(\delta\cdot\Pi_{\min})^{m+k}}$,
and thus,
$\rho_{j}\geq(1+\frac{M-1}{(\delta\cdot\Pi_{\min})^{m+k}})^{-1}$,
which proves the lemma.

\begin{lem}\label{second lemma for uniform convergence}
\emph{Consider following two arbitrarily given sets.
\begin{eqnarray*}
 C_{t}&\in&\mathcal{X}_{t}^{\infty}\triangleq \Big\{x_{\mathcal{T}}:\mathcal{T}\subseteq\mathbb{Z}_{\geq t}\cup\{\infty\} \Big\}\quad\textrm{and}\\
 D&\in&\mathcal{Z}_{-\infty}^{\infty}\triangleq \Big\{z_{\mathcal{T}}:\mathcal{T}\subseteq\mathbb{Z}\cup\{\infty,-\infty\} \Big\}.
\end{eqnarray*}
For $d\in\mathbb{N}$, define
\begin{eqnarray*}
 M_d^{+}&\triangleq&\max_{i}Q(C_{t}|X_{t-dm}=i,D),\\
 M_d^{-}&\triangleq&\min_{i}Q(C_{t}|X_{t-dm}=i,D).
\end{eqnarray*}
Then,
\begin{displaymath}
M_d^{+}-M_d^{-}\leq(\rho_{\delta,k,m})^{d-1}
\end{displaymath}
where $\rho_{\delta,k,m}=1-2\mu_{\delta,k,m}$.}
\end{lem}
\emph{Proof:} From the argument of Lemma \ref{first lemma for
uniform convergence}, it is easy to see that
$$
Q(X_{t+m}=j|X_{t}=i,D)\geq\mu_{\delta,k,m},
$$
independent of $D$, too. Now, define
\begin{eqnarray*}
\gamma_{i}(d)&\triangleq& Q(C_{t}|X_{t-dm}=i,D)\\
\beta_{ij}(d)&\triangleq& Q(X_{t-dm}=j|X_{t-(d+1)m}=i,D)\\
i^{+}(d)&\triangleq& \arg\max_{i}Q(C_{t}|X_{t-(d+1)m}=i,D)\\
i^-(d)&\triangleq& \arg\min_{i}Q(C_{t}|X_{t-dm}=i,D).
\end{eqnarray*}
Since $\delta$,$k$ and $m$ are fixed, let's simply denote
$\mu=\mu_{\delta,k,m}$. Also, let's omit $d$ and the parenthesis for
above four quantities to simplify notation. Then,
\begin{align}
M_{d+1}^{+}=&Q(C_{t}|X_{t-(d+1)m}=i^+,D)=\sum_{j}\gamma_{j}\beta_{i^+j}\nonumber\\
           =&\mu M_{d}^{-}+(\beta_{i^+i^-}-\mu)M_{d}^{-}+\sum_{j\neq
           i^-}\gamma_{j}\beta_{i^+j}\label{l7 eq1}\\
           \leq&\mu M_{d}^{-}+(\beta_{i^+i^-}-\mu)M_{d}^{+}+\sum_{j\neq
           i^-}\beta_{i^+j}M_{d}^{+}\nonumber\\
           =&\mu M_{d}^{-}+(1-\mu)M_{d}^{+}\label{l7 eq2}
\end{align}
where \eq{l7 eq1} is possible from Lemma \ref{first lemma for
uniform convergence}, since $\beta_{ij}\geq\mu$ for $\forall
i,j$.\\By the similar argument, we get
\begin{align}
M_{d+1}^{-}\geq& \mu M_{d}^{+}+(1-\mu)M_{d}^{-}\label{l7 eq3}
\end{align}
By subtracting \eq{l7 eq3} from \eq{l7 eq2}, we get
\begin{displaymath}
M_{d+1}^{+}-M_{d+1}^{-}\leq(1-2\mu)(M_{d}^{+}-M_{d}^{-})\leq\cdots\leq(1-2\mu)^{d}
\end{displaymath}
and, thus proves the lemma. Note that since
$\mu=\mu_{\delta,k,m}<\frac{1}{2}$, and thus,
$0<\rho_{\delta,k,m}<1$. Also, the result does not depend on $Q$.

\begin{lem}\label{third lemma for uniform convergence}
\emph{$$
|Q(C_{t}|Z_{t-dm-l}^{p})-Q(C_{t}|Z_{t-(d+1)m-l}^{p})|\leq(\rho_{\delta,k,m})^{d+1}
$$
for $\forall p$, $\forall d\geq 1$, and $0\leq l\leq m-1$.}
\end{lem}

\emph{Proof:}
\begin{align*}
 &Q(C_{t}|Z_{t-(d+1)m-l}^{p})\\
 =&\sum_{j}Q(C_{t}|Z_{t-(d+1)m-l}^{p},X_{t-(d+2)m}=j)Q(X_{t-(d+2)m}=j|Z_{t-(d+1)m-l}^{p})
\end{align*}
and therefore,
\begin{displaymath} M_{d+2}^{-}\leq
Q(C_{t}|Z_{t-(d+1)m-l}^{p})\leq M_{d+2}^{+}
\end{displaymath}
On the other hand,
\begin{align*}
 &Q(C_{t}|Z_{t-dm-l}^{p})\\
 =&\sum_{z_{t-(d+1)m-l}^{t-dm-l-1}}Q(C_{t}|Z_{t-(d+1)m-l}^{p})Q(Z_{t-(d+1)m-l}^{t-dm-l-1}=z_{t-(d+1)m-l}^{t-dm-l-1}|Z_{t-dm-l}^{p})
\end{align*}
and thus,
\begin{displaymath}
M_{d+2}^{-}\leq Q(C_{t}|Z_{t-dm-l}^{p})\leq M_{d+2}^{+}
\end{displaymath}
Therefore, from Lemma \ref{second lemma for uniform convergence}, we
have
$$
|Q(C_{t}|Z_{t-dm-l}^{p})-Q(C_{t}|Z_{t-(d+1)m-l}^{p})|\leq
M_{d+2}^{+}-M_{d+2}^{-} \leq(\rho_{\delta,k,m})^{d+1}
$$
Note that the result does not depend on either $Q$ or $l$.

\section*{Appendix 2}\label{Proof of lemma for
concentration} Before proving Lemma \ref{lemma for concentration} we
need following lemma first. Part (b),(c), and (d) are crucial for
Lemma \ref{lemma for concentration}, and Part (a) enables Part(b).
Part (a) is the reason why we need a randomization of the filter.

\begin{lem}\label{lemma for lemma for concentration}
\emph{Suppose $Q\in\Theta_k^{\delta}$ and fix
$\delta>0$.
\begin{itemize}
\item [(a)]
We have
$$
\|\hat{X}_Q^{\epsilon}(z_{-t_1}^{0})-\hat{X}_Q^{\epsilon}(z_{-t_2}^{0})\|_1\leq
M^2\cdot\|\mathbf{Q}_{X_0|z_{-t_1}^{0}}-\mathbf{Q}_{X_0|z_{-t_2}^{0}}\|_1,
$$
where $t_1,t_2>0$ are arbitrary integers. That is, for any integer
$t>0$ and any individual sequence $z_{-t}^{0}$,
$\hat{X}_Q^{\epsilon}(z_{-t}^{0})$ is a Lipschitz continuous
function in $\mathbf{Q}_{X_0|z_{-t}^{0}}$.
\item [(b)]
$\ell(X_0,\hat{X}_{Q}^{\epsilon}(Z_{-t}^{0}))\rightarrow\ell(X_0,\hat{X}_{Q}^{\epsilon}(Z_{-\infty}^{0}))$
\emph{a.s.} uniformly on $\Theta_{k}^{\delta}$
\item [(c)]
For $\forall Q\in\Theta_k^{\delta}$, and $\forall \omega$, $\exists
\quad 0<\gamma<1, \beta>0$, such that
$|Q(X_0|Z_{-t}^{0})-Q(X_0|Z_{-\infty}^{0})|<\beta\gamma^t$.
\item [(d)]
For fixed $t$,$\eta>0$, $\exists$ some finite set
$\mathcal{F}_k(t,\eta)\subset\Theta_k^{\delta}$, such that
$$
\max_{Q\in\Theta_{k}^{\delta}}\min_{Q'\in\mathcal{F}_k(t,\eta)}\max_{x_0,z_{-t}^{0}}|Q(x_0|z_{-t}^{0})-Q'(x_0|z_{-t}^{0})|\leq\eta
$$
\end{itemize}}
\end{lem}

\emph{Proof:}

\begin{itemize}
\item[(a)]
For given simplex vector $\mathbf{Q}$, fixed $\hat{x}$, and
$B_{\epsilon}$ defined as in Section \ref{sec: description of the
filter}, we define followings.
\begin{itemize}
\item[$\bullet$] $S_{\hat{x}}(\mathbf{Q})\triangleq\{\mathbf{W}\in
B_\epsilon:B(\mathbf{Q}+\mathbf{W})=\hat{x}\}$
\item[$\bullet$]
$DP(\hat{x})\triangleq\Big\{\mathbf{c}^T\mathbf{y}=0:\mathbf{y}\in\mathbb{R}^M,\mathbf{c}=\lambda_{\hat{x}}-\lambda_a,
\forall a\in\mathcal{A}\backslash\{\hat{x}\}\Big\}$
\item[$\bullet$] $\textrm{dist}(\mathbf{Q},\mathbf{c}^T\mathbf{y}=0)$ $\triangleq$ The shortest $L_2$
distance from a simplex vector $\mathbf{Q}$ to the plane
$\mathbf{c}^T\mathbf{y}=0$
\end{itemize}
That is, $S_{\hat{x}}(\mathbf{Q})$ is a set of vectors in
$\epsilon$-ball, $B_{\epsilon}$, that makes the Bayes response
$B(\mathbf{Q}+\mathbf{W})$ equal to $\hat{x}$. Also, $DP(\hat{x})$
is a set of decision planes that separate the decision region for
the reconstruction alphabet $\hat{x}$ and other alphabets. Then, for
some fixed $t$, by definition,
$$
\hat{X}_Q^{\epsilon}(z^0_{-t})[\hat{x}]=\frac{\textrm{Vol}(S_{\hat{x}}(\mathbf{Q}_{X_0|z_{-t}^{0}}))}{\textrm{Vol}(B_\epsilon)},
$$
where $\textrm{Vol}(\cdot)$ is a volume of a set.
Since $\textrm{Vol}(B_\epsilon)$ is a constant, for any $t_1$ and
$t_2$, we have
\begin{align}
|\hat{X}_Q^{\epsilon}(z^0_{-t_1})[\hat{x}]-\hat{X}_Q^{\epsilon}(z^0_{-t_2})[\hat{x}]|=\frac{|\textrm{Vol}(S_{\hat{x}}(\mathbf{Q}_{X_0|z_{-t_1}^{0}}))-\textrm{Vol}(S_{\hat{x}}(\mathbf{Q}_{X_0|z_{-t_2}^{0}}))|}{\textrm{Vol}(B_\epsilon)}.\label{append1
eq2}
\end{align}
For the numerater, as a crude bound, we get
\begin{align}
 &|\textrm{Vol}(S_{\hat{x}}(\mathbf{Q}_{X_0|z_{-t_1}^{0}}))-\textrm{Vol}(S_{\hat{x}}(\mathbf{Q}_{X_0|z_{-t_2}^{0}}))|\nonumber\\
 \leq&\textrm{Vol}(B_\epsilon^{M-1})\cdot\sum_{\mathbf{c}^T\mathbf{y}=0\in
DP(\hat{x})}\Big|\textrm{dist}(\mathbf{Q}_{X_0|z_{-t_1}^{0}},\mathbf{c}^T\mathbf{y}=0)-\textrm{dist}(\mathbf{Q}_{X_0|z_{-t_2}^{0}},\mathbf{c}^T\mathbf{y}=0)\Big|,\label{append1
eq1}
\end{align}
where
$B_\epsilon^{M-1}=\{\mathbf{U}\in\mathbb{R}^{M-1}:\|\mathbf{U}\|_2\leq\epsilon\}$.
Since
$$
\textrm{dist}(\mathbf{Q},\mathbf{c}^T\mathbf{y}=0)=\frac{|\mathbf{c}^T\mathbf{Q}|}{\|\mathbf{c}\|_2},
$$
we have
\begin{align}
 &\textrm{dist}(\mathbf{Q}_{X_0|z_{-t_1}^{0}},\mathbf{c}^T\mathbf{y}=0)-\textrm{dist}(\mathbf{Q}_{X_0|z_{-t_2}^{0}},\mathbf{c}^T\mathbf{y}=0)\nonumber\\
 =&\frac{|\mathbf{c}^T\mathbf{Q}_{X_0|z_{-t_1}^{0}}|-|\mathbf{c}^T\mathbf{Q}_{X_0|z_{-t_2}^{0}}|}{\|\mathbf{c}\|_2}\nonumber\\
\leq&\frac{\Big|\mathbf{c}^T(\mathbf{Q}_{X_0|z_{-t_1}^{0}}-\mathbf{Q}_{X_0|z_{-t_2}^{0}})\Big|}{\|\mathbf{c}\|_2}\label{append1 ineq1}\\
\leq&\|\mathbf{Q}_{X_0|z_{-t_1}^{0}}-\mathbf{Q}_{X_0|z_{-t_2}^{0}}\|_2\label{append1 ineq2}\\
\leq&\|\mathbf{Q}_{X_0|z_{-t_1}^{0}}-\mathbf{Q}_{X_0|z_{-t_2}^{0}}\|_1\label{append1
ineq3}
\end{align}
where \eq{append1 ineq1} is from the triangular inequality,
\eq{append1 ineq2} is from Cauchy-Schwartz inequality, and
\eq{append1 ineq3} is from the fact that $L_2$-norm is less than or
equal to $L_1$-norm. Therefore, \eq{append1 eq1} becomes
\begin{align*}
|\textrm{Vol}(S_{\hat{x}}(\mathbf{Q}_{X_0|z_{-t_1}^{0}})-\textrm{Vol}(S_{\hat{x}}(\mathbf{Q}_{X_0|z_{-t_2}^{0}})|\leq
M\cdot\textrm{Vol}(B_\epsilon^{M-1})\cdot\|\mathbf{Q}_{X_0|z_{-t_1}^{0}}-\mathbf{Q}_{X_0|z_{-t_2}^{0}}\|_1,
\end{align*}
and thus, \eq{append1 eq2} becomes
\begin{align*}
|\hat{X}_Q^{\epsilon}(z^0_{-t_1})[\hat{x}]-\hat{X}_Q^{\epsilon}(z^0_{-t_2})[\hat{x}]|\leq&M\cdot\frac{\textrm{Vol}(B_\epsilon^{M-1})}{\textrm{Vol}(B_\epsilon)}\cdot\|\mathbf{Q}(X_0|z_{-t_1}^{0})-\mathbf{Q}(X_0|z_{-t_2}^{0})\|_1.\\
\leq&M\cdot\|\mathbf{Q}_{X_0|z_{-t_1}^{0}}-\mathbf{Q}_{X_0|z_{-t_2}^{0}}\|_1.
\end{align*}
Therefore, we have
\begin{align*}
\|\hat{X}_Q^{\epsilon}(z_{-t_1}^{0})-\hat{X}_Q^{\epsilon}(z_{-t_2}^{0})\|_1\leq
M^2\cdot\|\mathbf{Q}_{X_0|z_{-t_1}^{0}}-\mathbf{Q}_{X_0|z_{-t_2}^{0}}\|_1,
\end{align*}
and Part (a) is proved.
\item[(b)] By the exact same argument as in proving Lemma 1, we can
easily know that $Q(X_0|Z_{-t}^{0})\rightarrow
Q(X_0|Z_{-\infty}^{0})$ for $\forall \omega$, uniformly in $\forall
Q\in\Theta_{k}^{\delta_k}$. Since we have
\begin{align*}
&\Big|\ell(X_0,\hat{X}_{Q}^{\epsilon}(Z_{-t}^{0}))-\ell(X_0,\hat{X}_Q^{\epsilon}(Z_{-\infty}^{0}))\Big|\\
=&\Big|\sum_{\hat{x}}\Lambda(X_0,\hat{x})\Big(\hat{X}_{Q}^{\epsilon}(Z_{-t}^{0})[\hat{x}]-\hat{X}_Q^{\epsilon}(Z_{-\infty}^{0})[\hat{x}]\Big)\Big|\\
\leq&\Lambda_{\max}\|\hat{X}_{Q}^{\epsilon}(Z_{-t}^{0})-\hat{X}_{Q}^{\epsilon}(Z_{-\infty}^{0})\|_{1}\\
\leq&\Lambda_{\max}
M^2\cdot\|\mathbf{Q}(X_0|Z_{-t}^{0})-\mathbf{Q}(X_0|Z_{-\infty}^{0})\|_{1},
\end{align*}
we get the uniform convergence.
\item[(c)] Again, let's follow the argument in the proof of Lemma 1.
Suppose $t=jk+l$, where $j=\lfloor t/k \rfloor$, and $l=t\mod k$.
Then,
\begin{align}
 &|Q(X_0|Z_{-t}^{0})-Q(X_0|Z_{-\infty}^{0})|\nonumber\\
=&|Q(X_0|Z_{-jk-l}^{0})-Q(X_0|Z_{-\infty}^{0})|\nonumber\\
\leq&\sum_{i=j}^{\infty}|Q(X_0|Z_{-ik-l}^{0})-Q(X_0|Z_{-(i+1)k-l}^{0})|\nonumber\\
\leq&\sum_{i=j}^{\infty}\rho^{i+1}\label{l9 eq1}\\
=&\frac{\rho^{j+1}}{1-\rho}=\frac{\rho}{1-\rho}\rho^{\lfloor t/k
\rfloor}=\frac{\rho^{1-\frac{l}{k}}}{1-\rho}(\rho^{1/k})^t\\
\leq&\frac{1}{1-\rho}(\rho^{1/k})^t
\end{align}
where $\rho=\rho_{\delta,k,k}$ as defined in Lemma 7, and \eq{l9
eq1} follows from Lemma 8. By letting $\beta=\frac{1}{1-\rho}$, and
$\gamma=\rho^{1/k}$, we have proved Part (c).
\item[(d)]
We know that for the individual sequence pair $(x_0,z_{-t}^{0})$,
\begin{align*}
Q(x_0|z_{-t}^{0})
=&\frac{\sum_{x_{-t}^{-1}}Q(x_{-t}^{0},z_{-t}^{0})}{Q(z_{-t}^{0})}\\
=&\frac{\sum_{x_{-t}^{-1}}Q(x_{-t}^{0},z_{-t}^{0})}{\sum_{x_{-t}^{0}}Q(x_{-t}^{0},z_{-t}^{0})}\\
=&\frac{\sum_{x_{-t}^{-1}}Q(x_{-t}^{0})Q(z_{-t}^{0}|x_{-t}^{0})}{\sum_{x_{-t}^{0}}Q(x_{-t}^{0})Q(z_{-t}^{0}|x_{-t}^{0})}\\
=&\frac{\sum_{x_{-t}^{-1}}\Big(Q(x_{-t}^{0})\prod_{i=-t}^{0}
\Pi(x_i,z_i)\Big)}{\sum_{x_{-t}^{0}}\Big(Q(x_{-t}^{0})\prod_{i=-t}^{0}
\Pi(x_i,z_i)\Big)}.
\end{align*}
For $Q\in\Theta_k^{\delta}$, $\mathbf{\Pi}$ is fixed and we can
think of $\prod_{i=-t}^{0} \Pi(x_i,z_i)$ as a constant for the
individual sequence pair $(x_{-t}^0,z_{-t}^0)$. Since
$$
Q(x_{-t}^0)=Q(x_{-t}^{k-1-t})\prod_{j=k-t}^{0}a_{x_{j-k}^{j-1}x_{j-k+1}^j},
$$
$Q(x_0|z_{-t}^{0})$ is the ratio of two finite order polynomials of
$\{a_{ij}\}$, and as $\Theta_k^{\delta}$ is closed and bounded,
$Q(x_0|z_{-t}^{0})$ is a uniformly continuous function of
$\{a_{ij}\}$. Therefore, for given $\eta$, $\exists \epsilon(\eta)$
such that $\|Q-Q^{'}\|_1<\epsilon(\eta)$ implies
$$
\max_{x_0,z_{-t}^{0}}|Q(x_0|z_{-t}^{0})-Q'(x_0|z_{-t}^{0})|\leq
\eta,
$$
since there are only finite number of possible $(x_0,z_{-t}^0)$
pairs. Also, since $\Theta_{k}^{\delta}$ is compact, we can always
find a finite set, $\mathcal{F}_k(t,\eta)$ that for any
$Q\in\Theta_k^{\delta}$, there exists at least one
$Q'\in\mathcal{F}_k(t,\eta)$, that satisfies
$\|Q-Q^{'}\|_1<\epsilon(\eta)$. Therefore, Part (d) is
proved.\end{itemize}

\emph{Proof of Lemma \ref{lemma for concentration}:} To prove Lemma
\ref{lemma for concentration}, first consider following limit.
\begin{align}
 &\lim_{n\rightarrow\infty}E\Big(L_{\hat{\mathbf{X}}_{Q}^{\epsilon}}(X^{n},Z^{n})\Big)\nonumber\\
=&\lim_{n\rightarrow\infty}\frac{1}{n}\sum_{t=1}^{n}E\Big(\ell(X_t,\hat{X}_{Q}^{\epsilon}(Z^t))\Big)\nonumber\\
=&\lim_{t\rightarrow\infty}E\Big(\ell(X_t,\hat{X}_{Q}^{\epsilon}(Z^t))\Big)\label{lem5 ineq1}\\
=&\lim_{t\rightarrow\infty}E\Big(\ell(X_0,\hat{X}_Q^{\epsilon}(Z_{-(t-1)}^0))\Big)\label{lem5 ineq2}\\
=&E\Big(\ell(X_{0},\hat{X}_{Q}^{\epsilon}(Z_{-\infty}^{0}))\Big)\textrm{
uniformly on $\Theta_{k}^{\delta}$}\label{lem5 ineq3},
\end{align}
where \eq{lem5 ineq1} is from Ces\'aro's mean convergence theorem,
\eq{lem5 ineq2} is from stationarity, and \eq{lem5 ineq3} is from
Lemma \ref{lemma for lemma for concentration}(b) and bounded
convergence theorem. Thus, to complete the proof, we need to show
that
\begin{eqnarray}
\lim_{n\rightarrow\infty}L_{\hat{\mathbf{X}}_{Q}^{\epsilon}}(X^{n},Z^{n})=
E\Big(\ell(X_{0},\hat{X}_{Q}^{\epsilon}(Z_{-\infty}^{0}))\Big)\quad\textrm{\emph{a.s.}\quad
uniformly on $\Theta_{k}^{\delta}$}\label{lem5 ineq4}
\end{eqnarray}
Now, let's show the pointwise convergence in \eq{lem5 ineq4} without
the uniformity by using ergodic theorem. For given $Q$, define
\begin{eqnarray*}
g_{t,Q}(X,Z)&\triangleq&\ell(X_0,\hat{X}_{Q}^{\epsilon}(Z_{-(t-1)}^{0}))\\
g_{Q}(X,Z)&\triangleq&\ell(X_0,\hat{X}_{Q}^{\epsilon}(Z_{-\infty}^{0}))
\end{eqnarray*}
and denote by $T$ the shift operator. Then, what we should prove
becomes
$$
\lim_{n\rightarrow\infty}\frac{1}{n}\sum_{t=1}^{n}g_{t,Q}(T^{t}(X,Z))=E\Big(g_{Q}(X,Z)\Big)
\quad\textrm{ \emph{a.s.}}
$$
while the ergodic theorem gives
$$
\lim_{n\rightarrow\infty}\frac{1}{n}\sum_{t=1}^{n}g_{Q}(T^{t}(X,Z))=E\Big(g_{Q}(X,Z)\Big)
\quad\textrm{ \emph{a.s.}}
$$
Observe that
\begin{align*}
 &\Big|\frac{1}{n}\sum_{t=1}^{n}g_{t,Q}(T^{t}(X,Z))-\frac{1}{n}\sum_{t=1}^{n}g_{Q}(T^{t}(X,Z))\Big|\\
\leq&\frac{1}{n}\sum_{t=1}^{n}\Big|g_{t,Q}(T^{t}(X,Z))-g_{Q}(T^{t}(X,Z))\Big|\\
=&\frac{1}{n}\sum_{t=1}^{n}\Big|\ell(X_t,\hat{X}_{Q}^{\epsilon}(Z_1^t))-\ell(X_t,\hat{X}_{Q}^{\epsilon}(Z_{-\infty}^t))\Big|.\\
\end{align*}
Since Lemma \ref{lemma for lemma for concentration}(c) holds for
$\forall \omega$, we can think that the lemma holds for all
individual sequence pair $(x_0,z^0_{-\infty})$. Thus, it holds for
all individual pair $(x_t,z^t_{-\infty})$, too, and we can conclude
that $Q(X_t|Z_1^t)\rightarrow Q(X_t|Z^t_{-\infty})$ for $\forall
\omega$ as $t\rightarrow\infty$. Hence, by exactly the same argument
as Lemma \ref{lemma for lemma for concentration}(a) and Lemma
\ref{lemma for lemma for concentration}(b), we conclude that
$\ell(X_t,\hat{X}_{Q}^{\epsilon}(Z_1^t))\rightarrow\ell(X_t,\hat{X}_{Q}^{\epsilon}(Z_{-\infty}^t))$
almost surely as $t\rightarrow\infty$. Now, by Ces\'aro's mean
convergence theorem , we obtain
$$
\frac{1}{n}\sum_{t=1}^{n}\Big|\ell(X_t,\hat{X}_Q^{\epsilon}(Z_1^t))-\ell(X_t,\hat{X}_Q^{\epsilon}(Z_{-\infty}^t))\Big|\rightarrow
0\quad\textrm{\emph{a.s.}}
$$
Therefore, we get
$$
L_{\hat{\mathbf{X}}_Q^{\epsilon}}(X^{n},Z^{n})\rightarrow
E\Big(\ell(X_{0},\hat{X}_{Q}^{\epsilon}(Z_{-\infty}^{0}))\Big)\quad\textrm{\emph{a.s.}}
$$
\indent Note that up to this point we cannot guarantee the
uniformity of the convergence, since the ergodic theorem only gives
the individual convergence for each $Q$. To show the uniformity of
the convergence in \eq{lem5 ineq4}, first define the following
quantity for some fixed integer $t\in[1,n-1]$,
$$
L_{\hat{\mathbf{X}}_{Q,t}^{\epsilon}}(X^n,Z^n)=\frac{1}{n}\Big(\sum_{i=1}^{t}\ell(X_i,\hat{X}_Q^{\epsilon}(Z^i))+\sum_{i=t+1}^{n}\ell(X_i,\hat{X}_Q^{\epsilon}(Z_{i-t}^{i}))\Big).
$$
From Lemma \ref{lemma for lemma for concentration}(d), for any
$Q\in\Theta_k^{\delta}$ and fixed $t,\eta>0$, we can pick some
$Q'\in\mathcal{F}_k(t,\eta)$ such that
$\|Q-Q^{'}\|_1<\epsilon(\eta)$, and thus,
$$
\max_{x_0,z_{-t}^{0}}|Q(x_0|z_{-t}^{0})-Q'(x_0|z_{-t}^{0})|\leq
\eta.
$$
By adding and subtracting some common terms involving such $Q^{'}$,
and from the triangle inequality, we have,
\begin{align}
 &\Big|L_{{\hat{\mathbf{X}}}_{Q}^{\epsilon}}(X^n,Z^n)-E\Big(\ell(X_{0},\hat{X}_{Q}^{\epsilon}(Z_{-\infty}^{0}))\Big)\Big|\nonumber\\
\leq&\Big|L_{{\hat{\mathbf{X}}}_{Q}^{\epsilon}}(X^n,Z^n)-L_{\hat{\mathbf{X}}_{Q,t}^{\epsilon}}(X^n,Z^n)\Big|+\Big|L_{\hat{\mathbf{X}}_{Q,t}^{\epsilon}}(X^n,Z^n)-L_{\hat{\mathbf{X}}_{Q',t}^{\epsilon}}(X^n,Z^n)\Big|+\Big|L_{\hat{\mathbf{X}}_{Q',t}^{\epsilon}}(X^n,Z^n)-L_{{\hat{\mathbf{X}}}_{Q'}^{\epsilon}}(X^n,Z^n)\Big|\nonumber\\
+&\Big|L_{{\hat{\mathbf{X}}}_{Q'}^{\epsilon}}(X^n,Z^n)-E\Big(\ell(X_0,\hat{X}_{Q'}^{\epsilon}(Z_{-\infty}^{0}))\Big)\Big|+\Big|E\Big(\ell(X_0,\hat{X}_{Q'}^{\epsilon}(Z_{-\infty}^{0}))\Big)-E\Big(\ell(X_0,\hat{X}_{Q}^{\epsilon}(Z_{-\infty}^{0}))\Big)\Big|\label{lem5
goal}
\end{align}
Now, the goal becomes to show that the terms in the righthand side
of the inequality converges to zero independent of $Q$ as $n$, $t$,
and $\eta$ varies. First, we will bound each term, and send
$n\rightarrow\infty$.
\begin{itemize}
\item [(1)]
\begin{align}
 &\Big|L_{{\hat{\mathbf{X}}}_{Q}^{\epsilon}}(X^n,Z^n)-L_{\hat{\mathbf{X}}_{Q,t}^{\epsilon}}(X^n,Z^n)\Big|\nonumber\\
\leq&\frac{1}{n}\sum_{i=t+1}^{n}\Big|\ell(X_i,\hat{X}_{Q}^{\epsilon}(Z^i))-\ell(X_i,\hat{X}_Q^{\epsilon}(Z_{i-t}^{i}))\Big|\nonumber\\
\leq&\Lambda_{\max}\cdot\frac{1}{n}\sum_{i=t+1}^{n}\|\hat{X}_Q^{\epsilon}(Z^i)-\hat{X}_Q^{\epsilon}(Z^{i}_{i-t})\|_1\nonumber\\
\leq&\Lambda_{\max}M^2\cdot\frac{1}{n}\sum_{i=t+1}^{n}\|\mathbf{Q}_{X_0|Z_{-i}^{0}}-\mathbf{Q}_{X_0|Z_{-t}^{0}}\|_1\label{lem5
ineq5}\\
\leq&\Lambda_{\max}M^3\cdot\frac{1}{n}\sum_{i=t+1}^{n}(\beta\gamma^t+\beta\gamma^i)\label{lem5
ineq6}\\
\rightarrow&\Lambda_{\max}M^3\beta\gamma^t\quad\textrm{\emph{a.s.}
 uniformly on $\Theta_k^{\delta}$}\label{lem5
ineq7}
\end{align}
where \eq{lem5 ineq5} is from stationarity and Lemma \ref{lemma for
lemma for concentration}(a), \eq{lem5 ineq6} is from Lemma
\ref{lemma for lemma for concentration}(c), and \eq{lem5 ineq7} is
from the Ces\'aro's mean convergence theorem. Since \eq{lem5 ineq6}
does not depend on $Q$, the limit is uniform on $\Theta_k^{\delta}$.
\item [(2)]
\begin{align}
 &\Big|L_{\hat{\mathbf{X}}_{Q,t}^{\epsilon}}(X^n,Z^n)-L_{\hat{\mathbf{X}}_{Q',t}^{\epsilon}}(X^n,Z^n)\Big|\nonumber\\
\leq&\frac{1}{n}\sum_{i=t+1}^{n}|\ell(X_i,\hat{X}_Q^{\epsilon}(Z^i_{i-t}))-\ell(X_i,\hat{X}_{Q'}^{\epsilon}(Z^i_{i-t}))|+\frac{t}{n}\cdot\Lambda_{max}\nonumber\\
\leq&\Lambda_{\max}\cdot\frac{1}{n}\sum_{i=t+1}^{n}\|\hat{X}_Q^{\epsilon}(Z^{i}_{i-t})-\hat{X}_{Q'}^{\epsilon}(Z^{i}_{i-t})\|_1+\frac{t}{n}\cdot\Lambda_{max}\nonumber\\
\leq&\Lambda_{\max}M^2\cdot\frac{1}{n}\sum_{i=t+1}^{n}\|\mathbf{Q}_{X_i|Z_{i-t}^{i}}-\mathbf{Q'}_{X_i|Z_{i-t}^{i}}\|_1+\frac{t}{n}\cdot\Lambda_{max}\label{lem5
ineq8}\\
\leq&\Lambda_{\max}M^3\frac{n-t}{n}\cdot\eta+\frac{t}{n}\cdot\Lambda_{max}\label{lem5
ineq9}\\
\rightarrow&\Lambda_{\max}M^3\eta\quad\textrm{\emph{a.s.} uniformly
on $\Theta_k^{\delta}$}\nonumber
\end{align}
where \eq{lem5 ineq8} is from Lemma \ref{lemma for lemma for
concentration}(a), and \eq{lem5 ineq9} is from Lemma \ref{lemma for
lemma for concentration}(d). Since \eq{lem5 ineq9} does not depend
on $Q$, the limit is also uniform on $\Theta_k^{\delta}$.
\item [(3)]
\begin{eqnarray*}
\Big|L_{\hat{\mathbf{X}}_{Q',t}^{\epsilon}}(X^n,Z^n)-L_{{\hat{\mathbf{X}}}_{Q'}^{\epsilon}}(X^n,Z^n)\Big|\rightarrow\Lambda_{max}M^3\beta\gamma^t\textrm{\quad
\emph{a.s.}}
\end{eqnarray*}
by following the same argument as (1). Since $\mathcal{F}_k(t,\eta)$
is finite, this convergence is uniform on $\mathcal{F}_k(t,\eta)$.
\item [(4)]
\begin{eqnarray*}
\Big|L_{{\hat{\mathbf{X}}}_{Q'}^{\epsilon}}(X^n,Z^n)-E\Big(\ell(X_0,\hat{X}_{Q'}^{\epsilon}(Z_{-\infty}^{0}))\Big)\Big|\rightarrow0\quad\textrm{\emph{a.s.}}
\end{eqnarray*}
from the proof of pointwise convergence above. As in (3), this
convergence is also uniform on $\mathcal{F}_k(t,\eta)$.
\item [(5)]
\begin{align*}
 &\Big|E\Big(\ell(X_0,\hat{X}_{Q'}^{\epsilon}(Z_{-\infty}^{0}))\Big)-E\Big(\ell(X_0,\hat{X}_{Q}^{\epsilon}(Z_{-\infty}^{0}))\Big)\Big|\\
\leq&\Big|E\Big(\ell(X_0,\hat{X}_{Q'}^{\epsilon}(Z_{-\infty}^{0}))\Big)-E\Big(\ell(X_0,\hat{X}_{Q'}^{\epsilon}(Z_{-t}^{0}))\Big)\Big|+\Big|E\Big(\ell(X_0,\hat{X}_{Q'}^{\epsilon}(Z_{-t}^{0}))\Big)-E[\ell(X_0,\hat{X}_{Q}^{\epsilon}(Z_{-t}^{0}))\Big)\Big|\\
+&\Big|E\Big(\ell(X_0,\hat{X}_{Q}^{\epsilon}(Z_{-t}^{0}))\Big)-E\Big(\ell(X_0,\hat{X}_{Q}^{\epsilon}(Z_{-\infty}^{0}))\Big)\Big|\\
\leq&\sum_{x_0,z_{-\infty}^{0}}P(x_0,z_{-\infty}^{0})\Big|\ell(x_0,\hat{X}_{Q'}^{\epsilon}(z_{-\infty}^{0}))-\ell(x_0,\hat{X}_{Q'}^{\epsilon}(z_{-t}^{0}))\Big|+\sum_{x_0,z_{-t}^{0}}P(x_0,z_{-t}^{0})\Big|\ell(x_0,\hat{X}_{Q'}^{\epsilon}(z_{-t}^{0}))-\ell(x_0,\hat{X}_{Q}^{\epsilon}(z_{-t}^{0}))\Big|\\
+&\sum_{x_0,z_{-\infty}^{0}}P(x_0,z_{-\infty}^{0})\Big|\ell(x_0,\hat{X}_{Q}^{\epsilon}(z_{-\infty}^{0}))-\ell(x_0,\hat{X}_{Q}^{\epsilon}(z_{-t}^{0}))\Big|\\
\leq&\Lambda_{max}M^3\Big(2\beta\gamma^t+\eta\Big),
\end{align*}
by similar argument as in (1) and (2).
\end{itemize}
Therefore, by taking limit supremum on both side of \eq{lem5 goal},
we get
\begin{align*}
 &\limsup_{n\rightarrow\infty}\Big|L_{{\hat{\mathbf{X}}}_{Q}^{\epsilon}}(X^n,Z^n)-E\Big(\ell(X_{0},\hat{X}_{Q}^{\epsilon}(Z_{-\infty}^{0}))\Big)\Big|\\
\leq&\Lambda_{max}M^3\Big(4\beta\gamma^t+2\eta\Big)\quad\textrm{\emph{a.s.}\quad
uniformly on $\Theta_k^{\delta}$}.
\end{align*}
Since $t$ and $\eta$ are arbitrary, by sending $t\rightarrow\infty$
and $\eta\downarrow0$, we have
$$
\limsup_{n\rightarrow\infty}\Big|L_{{\hat{\mathbf{X}}}_{Q}^{\epsilon}}(X^n,Z^n)-E\Big(\ell(X_{0},\hat{X}_{Q}^{\epsilon}(Z_{-\infty}^{0}))\Big)\Big|\leq0\quad\textrm{\emph{a.s.}\quad
uniformly on $\Theta_k^{\delta}$}.
$$
Therefore, the lemma is proved.\quad $\blacksquare$

\section*{Appendix 3}
Here, we prove Corollary \ref{univ concentration}.\\
\emph{Proof of Corollary \ref{univ concentration}:} First note the
subtle point that Corollary \ref{univ concentration} does not
directly follow from Lemma \ref{lemma for concentration}. Since the
probability law $Q_k^t$ that we are using to filter each block is
changing every block, whereas the uniform convergence in Lemma
\ref{lemma for concentration} is for the fixed
$Q\in\Theta_k^{\delta_k}$ for all $t$, it is not enough to guarantee
the Corollary. However,  since $Q_k^t$ remains the same within each
block, we can still use the result of Lemma \ref{lemma for
concentration} if the block length gets long enough. Keeping this in
mind, let's take a more careful look at each block. In the proof,
for the brevity of notation, let's denote
$$
\ell_t(Q)\triangleq\ell(X_t,\hat{X}_Q^{\epsilon}(Z^t)),
$$
since we are always dealing with the randomized filter, and there is
no possibility of confusion. Now, fix any $\delta>0$. Then, from
\eq{condition on sequence},
$$
\exists I,\quad\textrm{such that}\quad
\frac{m_{I-1}}{m_{I}}<\frac{\delta}{8\ell_{max}},
$$
and from Lemma \ref{lemma for concentration},
$$
\exists N,\quad\textrm{such that}\quad
\max_{Q\in\Theta_K^{\delta_k}}\left|L_{\hat{\mathbf{X}}_{Q}^{\epsilon}}(X^n,Z^n)-EL_{\hat{\mathbf{X}}_{Q}^{\epsilon}}(X^n,Z^n)\right|<\delta/4.
$$
Recalling the definition $i(t)\triangleq\max\{i:m_i\leq t\}$, we let
$I_0=\max(I,i(N)+1)$. Then, for any $n\geq m_{I_0}$, and
$m_{i(n)}\leq n< m_{i(n)+1}$,
\begin{align}
 &\left|L_{\hat{\mathbf{X}}_{univ,k}^{\epsilon}}(X^n,Z^n)-\hat{E}L_{\hat{\mathbf{X}}_{univ,k}^{\epsilon}}(X^n,Z^n)\right|\\
\leq\frac{1}{n}&\left|\sum_{t=1}^{m_{i(n)-1}}\Big(\ell_t(Q_k^t)-\hat{E}(\ell_t(Q_k^t))\Big)\right|+\frac{1}{n}\left|\sum_{t=m_{i(n)-1}+1}^{m_{i(n)}}\Big(\ell_t(\hat{Q}[Z^{m_{i(n)-1}}])-\hat{E}(\ell_t(\hat{Q}[Z^{m_{i(n)-1}}]))\Big)\right|\\
 +\frac{1}{n}&\left|\sum_{t=m_{i(n)}+1}^{n}\Big(\ell_t(\hat{Q}[Z^{m_{i(n)}}])-\hat{E}(\ell_t(\hat{Q}[Z^{m_{i(n)}}]))\Big)\right|.
\end{align}
Note that in the second and third term, $Q_k^t$ is fixed to
$\hat{Q}[Z^{m_{i(n)-1}}]$ and $\hat{Q}[Z^{m_{i(n)}}]$ from the
definition of our filter. Now, we can bound each term. For the first
term, since $n\geq m_{i(n)}\geq m_{I}$, we know that
$\frac{m_{i(n)-1}}{n}\leq\frac{m_{i(n)-1}}{m_{i(n)}}<\frac{\delta}{8\ell_{max}}$.
Therefore,
$$
\frac{1}{n}\left|\sum_{t=1}^{m_{i(n)-1}}\Big(\ell_t(Q_k^t)-\hat{E}(\ell_t(Q_k^t))\Big)\right|\leq\frac{\delta}{8\ell_{max}}\cdot\ell_{max}=\frac{\delta}{8}.
$$
For the second term, since $n\geq m_{i(n)}\geq N$,
\begin{align}
 &\frac{1}{n}\left|\sum_{t=m_{i(n)-1}+1}^{m_{i(n)}}\Big(\ell_t(\hat{Q}[Z^{m_{i(n)-1}}])-\hat{E}(\ell_t(\hat{Q}[Z^{m_{i(n)-1}}]))\Big)\right|\\
\leq&\frac{m_{i(n)}}{n}\frac{1}{m_{i(n)}}\left|\sum_{t=1}^{m_{i(n)}}\Big(\ell_t(\hat{Q}[Z^{m_{i(n)-1}}])-\hat{E}(\ell_t(\hat{Q}[Z^{m_{i(n)-1}}]))\Big)\right|+\frac{1}{n}\left|\sum_{t=1}^{m_{i(n)-1}}\Big(\ell_t(\hat{Q}[Z^{m_{i(n)-1}}])-\hat{E}(\ell_t(\hat{Q}[Z^{m_{i(n)-1}}]))\Big)\right|\\
\leq&\frac{\delta}{4}+\frac{\delta}{8\ell_{max}}\cdot\ell_{max}=\frac{3\delta}{8}
\end{align}
Finally, for the last term,
\begin{align}
 &\frac{1}{n}\left|\sum_{t=m_{i(n)}+1}^{n}\Big(\ell_t(\hat{Q}[Z^{m_{i(n)}}])-\hat{E}(\ell_t(\hat{Q}[Z^{m_{i(n)}}]))\Big)\right|\\
 \leq&\frac{1}{n}\left|\sum_{t=1}^{n}\Big(\ell_t(\hat{Q}[Z^{m_{i(n)}}])-\hat{E}(\ell_t(\hat{Q}[Z^{m_{i(n)}}]))\Big)\right|+\frac{1}{n}\left|\sum_{t=1}^{m_{i(n)}}\Big(\ell_t(\hat{Q}[Z^{m_{i(n)}}])-\hat{E}(\ell_t(\hat{Q}[Z^{m_{i(n)}}]))\Big)\right|\\
 \leq&\frac{\delta}{4}+\frac{\delta}{4}=\frac{\delta}{2}.
\end{align}
Therefore, for any $n\geq m_{I_0}$, and $m_{i(n)}\leq n\leq
m_{i(n)+1}$, we have
$$
\left|L_{\hat{\mathbf{X}}_{univ,k}^{\epsilon}}(X^n,Z^n)-\hat{E}L_{\hat{\mathbf{X}}_{univ,k}^{\epsilon}}(X^n,Z^n)\right|<\delta,
$$
and since $\delta$ was arbitrary, we have the corollary.\quad
$\blacksquare$


\begin{thebibliography}{1}


\bibitem {BaumPetrie66}
\newblock L.E. Baum and T. Petrie, ``Statistical Inference for probabilistic
    functions of finite state Markov chains,''
\newblock {\em Ann. Math. Statist.}, vol.37, 1554-1563, 1966

\bibitem{BPSW70}
L.E. Baum, T.Petrie, G. Soules, and N. Weiss, \newblock ``A
maximization technique occuring in the statistical analysis of
probabilistic functions of Markov chains,'' \newblock {\em Ann.
Math. Statist.}, 41:164-171, 1970

\bibitem{BRR98}
P.J.~Bickel, Y.~Ritov, and T.~Ryd\'en, \newblock ``Asymptotic
normality of the maxumum-likelihood estimator for general hidden
Markov models,'' \newblock {\em Ann. Statist.}, 26(4):1614-1635,
1998

\bibitem{ChangHancock66}
R.W. ~Chang and J.C.~Hancock, \newblock ``On receiver structures for
channels having memory,'' \newblock {\em IEEE Trans. Inform.
Theory}, vol.IT-12:463-468, Oct 1996

\bibitem{CoverThomas91}
T.M. Cover and J.A. Thomas,
\newblock ``Elements of Information Theory,''
\newblock {\em New York: Wiley, 1991}

\bibitem{DemboWeissmancontoutputdude2003}
A.~Dembo and T.~Weissman,
\newblock ``Universal denoising for the finite-input-general-output
channel,''
\newblock {\em IEEE Trans. Inform. Theory}, 51(4):1507-1517, April 2005


\bibitem{EphraimMerhav2002}
Y.~Ephraim and N.~Merhav,
\newblock ``Hidden Markov processes,''
\newblock {\em IEEE Trans. Inform. Theory}, 48(6):1518-1569, June 2002


\bibitem{Finesso90}
L. Finesso,
\newblock ``Consistent Estimation of the Order for Markov and Hidden
    Markov Chains,'' \textit{Ph.D Dissertation, Univ. Maryland, College Park},
    1990

\bibitem{gemelos04}
G.~Gemelos, S.~Sigurj\'onsson, and T. ~Weissman,
\newblock ``Universal discrete denoising under channel
uncertainty,'' \newblock {\it  Proceedings of Int.\ Symp.\ Inf.\
Th.\ }, pp.199, Chicago, IL, Jun./Jul. 2004

\bibitem{GolubMeyer86}
G.H. Golub and C.D. Meyer, ``Using the QR factorization and group
inversion to compute, differentiate, and estimate the sensitivity of
stationary probabilities for Markov chains'' \textit{SIAM J.
Algebraic Discrete Meth.}, 7, pp.273-281, 1986

\bibitem{Kieffer93}
J.C. Kieffer, \newblock ``Strongly consistent code-based
identification and order estimation for constrained finite-state
model classes,''\newblock {\em IEEE Trans. Inform. Theory},
39:893-902, May 1993

\bibitem{Leroux92}
B.G. Leroux, ``Maximum-likelihood estimation for hidden Markov
models,'' \textit{Stochastic Processes Their Appic.}, vol. 40,
pp.127-143, 1992

\bibitem{LN94}
C.-C Liu and P. Narayan, \newblock ``Order estimation and sequential
universal data compression of a hidden Markov source bu the model of
mixtures,'' \newblock {\em IEEE Trans. Inform. Theory},40:1167-1180,
July 1994

\bibitem{Merhav1998}
N. Merhav and M. Feder, ``Universal Prediction,'' \textit{IEEE
Trans. Inform. Theory}, 44(6):2124-2147, Oct 1998

\bibitem{MerhavGutmanFeder92}
N. Feder, N. Merhav, and M. Gutman, ``Universal Prediction for
individual sequences'' \textit{IEEE Trans. Inform. Theory},
38:1258-1270, July 1992


\bibitem{MevelFinesso04}
L. Mevel and L. Finesso,
\newblock ``Asymptotical statistics of misspecified hidden Markov
models,''
\newblock {\em IEEE Trans. Automatic Control}, 49(7):1123-1132, Jul
2004


\bibitem{Taesup05}
T. Moon and T. Weissman,
\newblock ``Discrete universal filtering via hidden Markov
modelling,''
\newblock {\it  Int.\ Symp.\ Inf.\ Th.\ },  Adelaide,
Australia, September 2005.

\bibitem{sequentialdude04}
E.~Ordentlich, T.~Weissman, M.~Weinberger, A.~Somekh-Baruch and
N.~Merhav,
\newblock ``Discrete universal filtering through incremental
parsing''
\newblock {\it Data Compression Conference (DCC 2004)}, p.\
352-361,
 Snowbird, Utah
 March 23-25, 2004.

\bibitem{filteringbyprediction}
E.~Ordentlich, T.~Weissman, M.~Weinberger, A.~Somekh-Baruch and
N.~Merhav,
\newblock ``Universal filtering via prediction'', available at
\verb"http://www.stanford.edu/~tsachy/filtering.pdf"

\bibitem{Ryden95}
T. Ryd\'en, \newblock ``Estimating the order of hidden Markov
models,'' \newblock {\em Statistics}, 26:345-354, 1995

\bibitem{DUDE}
T.~Weissman, E.~Ordentlich, G.~Seroussi, S.~Verd\'{u}, and
M.~Weinberger,
\newblock ``Universal discrete denoising: Known channel,''
\newblock {\em IEEE Trans. Inform. Theory}, 51(1):5-28, January 2005.

\bibitem{zhang05}
R. ~Zhang and T. ~Weissman, \newblock ``Discrete Denoising for
Channels with Memory'', {\em Communications in Informations and
Systems}, 5(2):257-288, 2005

\bibitem{ZivMerhav92}
J. Ziv and N. Merhav,\newblock ``Estimating the number of states of
a finite-state source,'' \newblock \newblock {\em IEEE Trans.
Inform. Theory},38:61-65, Jan 1992



\end{thebibliography}
\end{document}